\documentclass[]{llncs}

\usepackage[utf8]{inputenc}      % UTF-8 input encoding
\usepackage[T1]{fontenc}         % Type1 fonts
\usepackage{microtype}       % Better handling of typo
\usepackage[english]{babel}      % Hyphenation
\usepackage{graphicx}            % Import graphics
\usepackage{cite}                % Better handling of citations
\usepackage[scaled]{helvet} % Scale Helvetica
\usepackage[]{hyperref}            % Better handling of references in PDFs
\usepackage{comment}             % To comment out blocks of text
\usepackage{lmodern}             % Improved Computer Modern font

\title{Efficient Offline Monitoring of Linear Temporal Logic with Bit Vectors}

\author{Kun Xie \and Sylvain Hallé%
}
\institute{%
Laboratoire d'informatique formelle \\
Université du Québec à Chicoutimi, Canada%
}

\usepackage{amsmath,amsfonts,amssymb}

\usepackage{subfig}

\usepackage{algorithm}% http://ctan.org/pkg/algorithms
\usepackage{algpseudocode}% http://ctan.org/pkg/algorithmicx
\usepackage{multicol,multirow}
\algrenewcommand\algorithmicindent{0.5em}

\usepackage{cite}

\usepackage{xcolor}
\usepackage{todonotes}

\newcommand{\tG}{\ensuremath{\mbox{\textbf{G}}}\,}
\newcommand{\tF}{\ensuremath{\mbox{\textbf{F}}}\,}
\newcommand{\tX}{\ensuremath{\mbox{\textbf{X}}}\,}
\newcommand{\tU}{\ensuremath{\mbox{\,\textbf{U}}}\,}

\newcommand{\etal}{et al.}
\usepackage{longtable}

\graphicspath{{fig/}}

\begin{document}

\maketitle
\begin{abstract}
%% ----------------------
%% Write your abstract here. Do not enclose it in an "abstract"
%% environment.
%% ----------------------
A bitmap is a data structure designed to compactly represent sets of integers; it provides very fast operations for querying and manipulating such sets, exploiting bit-level parallelism. In this paper, we describe a technique for the offline verification of arbitrary expressions of Linear Temporal Logic using bitmap manipulation. An event trace is first preprocessed and transformed into a set of bitmaps. The LTL expression is then evaluated through a recursive procedure manipulating these bitmaps. Experimental results show that, for complex LTL formul\ae{} containing almost 20 operators, event traces can be evaluated at a throughput of millions of events per second.
%% :folding=explicit:wrap=soft:mode=latex:

\end{abstract}

%\input{preamble-stvr.inc.tex}
%\input{preamble-svjour.inc.tex}

%% ---------------------------
%% If you wish to include additional packages, define new environments or
%% new commands, put them in the file includes.tex
%%
%% Write your abstract in the file abstract.tex.
%% ---------------------------

%% ---------------------------
%% LabPal stuff
%% ---------------------------
% ----------------------------------------------------------------
% File generated by LabPal 2.11.6
% Date:     27-01-2020
% Lab name: Untitled
%
% To insert one of the figures into your text, do:
% \begin{figure}
% \usebox{\boxname}
% \end{figure}
% where \boxname is one of the boxes defined in the file below
% ----------------------------------------------------------------

% ----------------------
% Plot: pTimeMFormulaSZOStringPlengthO
% ----------------------
\newsavebox{\pTimeMFormulaSZOStringPlengthO}
\begin{lrbox}{\pTimeMFormulaSZOStringPlengthO}
\href{P1.0}{\includegraphics[page=1,width=\linewidth]{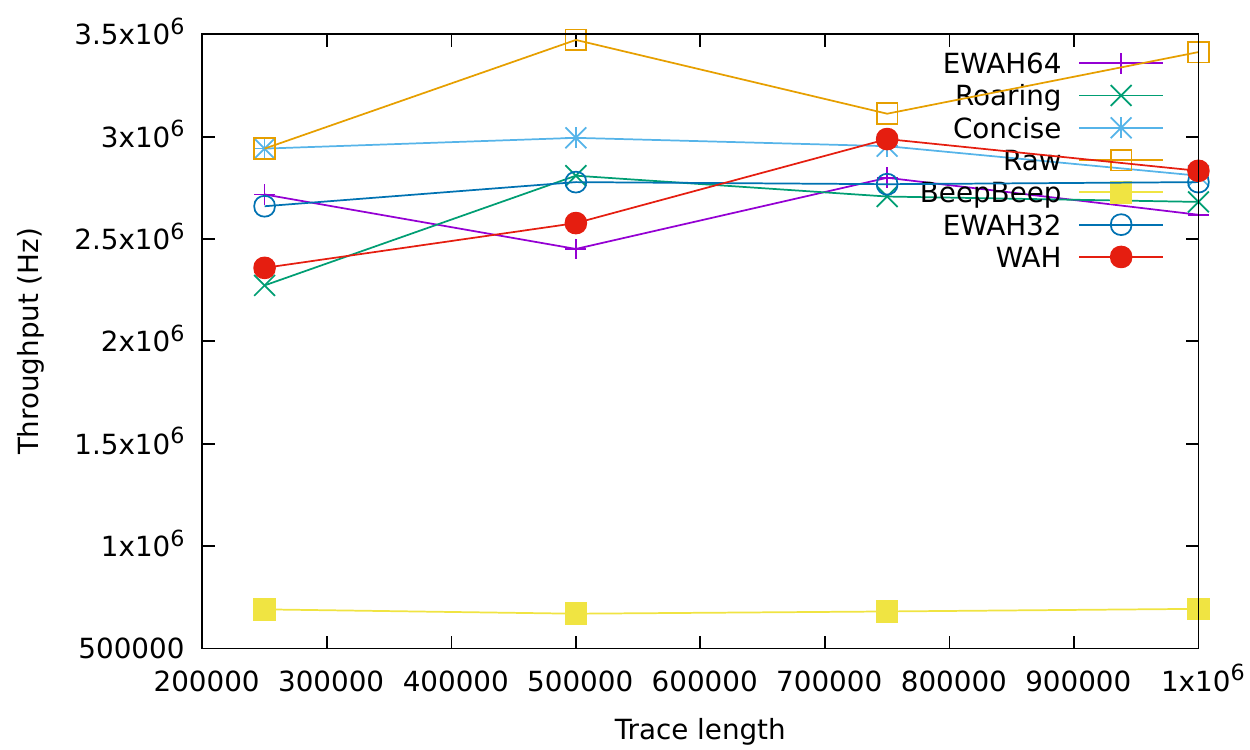}}
\end{lrbox}

% ----------------------
% Plot: pMemMFormulaSZOStringPlengthO
% ----------------------
\newsavebox{\pMemMFormulaSZOStringPlengthO}
\begin{lrbox}{\pMemMFormulaSZOStringPlengthO}
\href{P2.0}{\includegraphics[page=2,width=\linewidth]{labpal-plots.pdf}}
\end{lrbox}

% ----------------------
% Plot: pTimeMFormulaSZWStringPlengthO
% ----------------------
\newsavebox{\pTimeMFormulaSZWStringPlengthO}
\begin{lrbox}{\pTimeMFormulaSZWStringPlengthO}
\href{P3.0}{\includegraphics[page=3,width=\linewidth]{labpal-plots.pdf}}
\end{lrbox}

% ----------------------
% Plot: pMemMFormulaSZWStringPlengthO
% ----------------------
\newsavebox{\pMemMFormulaSZWStringPlengthO}
\begin{lrbox}{\pMemMFormulaSZWStringPlengthO}
\href{P4.0}{\includegraphics[page=4,width=\linewidth]{labpal-plots.pdf}}
\end{lrbox}

% ----------------------
% Plot: hThroughputLongestM
% ----------------------
\newsavebox{\hThroughputLongestM}
\begin{lrbox}{\hThroughputLongestM}
\href{P5.0}{\includegraphics[page=5,width=\linewidth]{labpal-plots.pdf}}
\end{lrbox}

% ----------------------
% Plot: bxRpTime
% ----------------------
\newsavebox{\bxRpTime}
\begin{lrbox}{\bxRpTime}
\href{P6.0}{\includegraphics[page=6,width=\linewidth]{labpal-plots.pdf}}
\end{lrbox}

% ----------------------
% Plot: hMemLongestM
% ----------------------
\newsavebox{\hMemLongestM}
\begin{lrbox}{\hMemLongestM}
\href{P7.0}{\includegraphics[page=7,width=\linewidth]{labpal-plots.pdf}}
\end{lrbox}

% ----------------------
% Plot: bxRpMem
% ----------------------
\newsavebox{\bxRpMem}
\begin{lrbox}{\bxRpMem}
\href{P8.0}{\includegraphics[page=8,width=\linewidth]{labpal-plots.pdf}}
\end{lrbox}

% ----------------------
% Plot: pThroughputVsSizeLongestConcise
% ----------------------
\newsavebox{\pThroughputVsSizeLongestConcise}
\begin{lrbox}{\pThroughputVsSizeLongestConcise}
\href{P9.0}{\includegraphics[page=9,width=\linewidth]{labpal-plots.pdf}}
\end{lrbox}

% ----------------------
% Plot: pThroughputVsSizeLongestEWAHRW
% ----------------------
\newsavebox{\pThroughputVsSizeLongestEWAHRW}
\begin{lrbox}{\pThroughputVsSizeLongestEWAHRW}
\href{P10.0}{\includegraphics[page=10,width=\linewidth]{labpal-plots.pdf}}
\end{lrbox}

% ----------------------
% Plot: pThroughputVsSizeLongestEWAHXF
% ----------------------
\newsavebox{\pThroughputVsSizeLongestEWAHXF}
\begin{lrbox}{\pThroughputVsSizeLongestEWAHXF}
\href{P11.0}{\includegraphics[page=11,width=\linewidth]{labpal-plots.pdf}}
\end{lrbox}

% ----------------------
% Plot: pThroughputVsSizeLongestRaw
% ----------------------
\newsavebox{\pThroughputVsSizeLongestRaw}
\begin{lrbox}{\pThroughputVsSizeLongestRaw}
\href{P12.0}{\includegraphics[page=12,width=\linewidth]{labpal-plots.pdf}}
\end{lrbox}

% ----------------------
% Plot: pThroughputVsSizeLongestRoaring
% ----------------------
\newsavebox{\pThroughputVsSizeLongestRoaring}
\begin{lrbox}{\pThroughputVsSizeLongestRoaring}
\href{P13.0}{\includegraphics[page=13,width=\linewidth]{labpal-plots.pdf}}
\end{lrbox}

% ----------------------
% Plot: pThroughputVsSizeLongestWAH
% ----------------------
\newsavebox{\pThroughputVsSizeLongestWAH}
\begin{lrbox}{\pThroughputVsSizeLongestWAH}
\href{P14.0}{\includegraphics[page=14,width=\linewidth]{labpal-plots.pdf}}
\end{lrbox}

% ----------------------
% Plot: pThroughputVsSizeLongestBeepBeep
% ----------------------
\newsavebox{\pThroughputVsSizeLongestBeepBeep}
\begin{lrbox}{\pThroughputVsSizeLongestBeepBeep}
\href{P15.0}{\includegraphics[page=15,width=\linewidth]{labpal-plots.pdf}}
\end{lrbox}

% ----------------------
% Plot: pTimeTSlenFormulaSZOTracePlengthOZZZZZZ
% ----------------------
\newsavebox{\pTimeTSlenFormulaSZOTracePlengthOZZZZZZ}
\begin{lrbox}{\pTimeTSlenFormulaSZOTracePlengthOZZZZZZ}
\href{P16.0}{\includegraphics[page=16,width=\linewidth]{labpal-plots.pdf}}
\end{lrbox}

% ----------------------
% Plot: pCompressionTSlenFormulaSZOTracePlengthOZZZZZZ
% ----------------------
\newsavebox{\pCompressionTSlenFormulaSZOTracePlengthOZZZZZZ}
\begin{lrbox}{\pCompressionTSlenFormulaSZOTracePlengthOZZZZZZ}
\href{P17.0}{\includegraphics[page=17,width=\linewidth]{labpal-plots.pdf}}
\end{lrbox}

% ----------------------
% Plot: pTimeTSlenFormulaSOOTracePlengthOZZZZZZ
% ----------------------
\newsavebox{\pTimeTSlenFormulaSOOTracePlengthOZZZZZZ}
\begin{lrbox}{\pTimeTSlenFormulaSOOTracePlengthOZZZZZZ}
\href{P18.0}{\includegraphics[page=18,width=\linewidth]{labpal-plots.pdf}}
\end{lrbox}

% ----------------------
% Plot: pCompressionTSlenFormulaSOOTracePlengthOZZZZZZ
% ----------------------
\newsavebox{\pCompressionTSlenFormulaSOOTracePlengthOZZZZZZ}
\begin{lrbox}{\pCompressionTSlenFormulaSOOTracePlengthOZZZZZZ}
\href{P19.0}{\includegraphics[page=19,width=\linewidth]{labpal-plots.pdf}}
\end{lrbox}

% ----------------------------------------------------------------
% File generated by LabPal 2.11.6
% Date:     27-01-2020
% Lab name: Untitled
% ----------------------------------------------------------------

% machinestring
% Basic info about the machine running the lab
\newcommand{\machinestring}{\href{M1.0}{Intel Xeon quad-core 2.66 GHz running Ubuntu 18.04}}

% machineram
% Total memory in the machine running the lab
\newcommand{\machineram}{\href{M1.1}{16 GB}}

% jvmram
% RAM available to the JVM
\newcommand{\jvmram}{\href{M1.2}{1746}}

% numexperiments
% The number of experiments in the lab
\newcommand{\numexperiments}{\href{M1.3}{465}}

% numdatapoints
% The number of data points in the lab
\newcommand{\numdatapoints}{\href{M1.4}{4398}}

% maxtracelen
% The maximum length of the traces
\newcommand{\maxtracelen}{\href{M2.0}{1000000}}

% numformulas
% The number of distinct LTL formulas tested
\newcommand{\numformulas}{\href{M2.1}{57}}

% nummethods
% The number of distinct compression methods tested
\newcommand{\nummethods}{\href{M2.2}{6}}

% minspeedup
% The minimum speedup provided by the best compression library, compared to BeepBeep
\newcommand{\minspeedup}{\href{M2.3}{1.1}}

% maxspeedup
% The maximum speedup provided by the best compression library, compared to BeepBeep
\newcommand{\maxspeedup}{\href{M2.4}{10.8}}

% winConcise
% The number of formulas for which method Concise had the highest throughput of all
\newcommand{\winConcise}{\href{M2.5}{1}}

% winEWAHRW
% The number of formulas for which method EWAH32 had the highest throughput of all
\newcommand{\winEWAHRW}{\href{M2.6}{null}}

% winEWAHXF
% The number of formulas for which method EWAH64 had the highest throughput of all
\newcommand{\winEWAHXF}{\href{M2.7}{null}}

% winRaw
% The number of formulas for which method Raw had the highest throughput of all
\newcommand{\winRaw}{\href{M2.8}{54}}

% winRoaring
% The number of formulas for which method Roaring had the highest throughput of all
\newcommand{\winRoaring}{\href{M2.9}{2}}

% winWAH
% The number of formulas for which method WAH had the highest throughput of all
\newcommand{\winWAH}{\href{M2.10}{null}}

% winBeepBeep
% The number of formulas for which method BeepBeep had the highest throughput of all
\newcommand{\winBeepBeep}{\href{M2.11}{null}}

% winner
% The technique that has the best throughput for the most formulas
\newcommand{\winner}{\href{M2.12}{Raw}}

% ----------------------------------------------------------------
% File generated by LabPal 2.11.6
% Date:     27-01-2020
% Lab name: Untitled
%
% To insert one of the tables into your text, do:
% \begin{table}
% \usebox{\boxname}
% \end{table}
% where \boxname is one of the boxes defined in the file below
% ----------------------------------------------------------------

% ----------------------
% Table: tTimeMFormulaSZOStringPlengthO
% ProcessingtimebymethodFormulaSabStringlengthb
% ----------------------
\newsavebox{\tTimeMFormulaSZOStringPlengthO}
\begin{lrbox}{\tTimeMFormulaSZOStringPlengthO}
\begin{tabular}{|c|c|c|c|c|c|c|c|}
\hline
\textbf{Trace length} & \textbf{EWAH64} & \textbf{Roaring} & \textbf{Concise} & \textbf{Raw} & \textbf{BeepBeep} & \textbf{EWAH32} & \textbf{WAH}\\
\hline\hline
{\href{T3.0.0}{250000}} & {\href{T3.0.1}{2717391.2}} & {\href{T3.0.2}{2272727.2}} & {\href{T3.0.3}{2941176.5}} & {\href{T3.0.4}{2941176.5}} & {\href{T3.0.5}{690607.75}} & {\href{T3.0.6}{2659574.5}} & {\href{T3.0.7}{2358490.5}}\\
\hline
{\href{T3.1.0}{500000}} & {\href{T3.1.1}{2450980.5}} & {\href{T3.1.2}{2808988.8}} & {\href{T3.1.3}{2994012.0}} & {\href{T3.1.4}{3472222.2}} & {\href{T3.1.5}{669344.06}} & {\href{T3.1.6}{2777777.8}} & {\href{T3.1.7}{2577319.5}}\\
\hline
{\href{T3.2.0}{750000}} & {\href{T3.2.1}{2798507.5}} & {\href{T3.2.2}{2707581.2}} & {\href{T3.2.3}{2952756.0}} & {\href{T3.2.4}{3112033.2}} & {\href{T3.2.5}{680580.75}} & {\href{T3.2.6}{2767527.8}} & {\href{T3.2.7}{2988047.8}}\\
\hline
{\href{T3.3.0}{1000000}} & {\href{T3.3.1}{2617801.0}} & {\href{T3.3.2}{2680965.2}} & {\href{T3.3.3}{2808988.8}} & {\href{T3.3.4}{3412969.2}} & {\href{T3.3.5}{693000.7}} & {\href{T3.3.6}{2777777.8}} & {\href{T3.3.7}{2832861.2}}\\

\hline
\end{tabular}
\end{lrbox}

% ----------------------
% Table: tMemMFormulaSZOStringPlengthO
% MemoryconsumptionbymethodFormulaSabStringlengthb
% ----------------------
\newsavebox{\tMemMFormulaSZOStringPlengthO}
\begin{lrbox}{\tMemMFormulaSZOStringPlengthO}
\begin{tabular}{|c|c|c|c|c|c|c|c|}
\hline
\textbf{Trace length} & \textbf{EWAH64} & \textbf{Roaring} & \textbf{Concise} & \textbf{Raw} & \textbf{BeepBeep} & \textbf{EWAH32} & \textbf{WAH}\\
\hline\hline
{\href{T4.0.0}{250000}} & {\href{T4.0.1}{406448}} & {\href{T4.0.2}{434424}} & {\href{T4.0.3}{419388}} & {\href{T4.0.4}{437500}} & {\href{T4.0.5}{14737}} & {\href{T4.0.6}{406336}} & {\href{T4.0.7}{419388}}\\
\hline
{\href{T4.1.0}{500000}} & {\href{T4.1.1}{812672}} & {\href{T4.1.2}{852368}} & {\href{T4.1.3}{838744}} & {\href{T4.1.4}{875000}} & {\href{T4.1.5}{14737}} & {\href{T4.1.6}{812560}} & {\href{T4.1.7}{838720}}\\
\hline
{\href{T4.2.0}{750000}} & {\href{T4.2.1}{1218896}} & {\href{T4.2.2}{1286680}} & {\href{T4.2.3}{1258096}} & {\href{T4.2.4}{1312500}} & {\href{T4.2.5}{14762}} & {\href{T4.2.6}{1218836}} & {\href{T4.2.7}{1258096}}\\
\hline
{\href{T4.3.0}{1000000}} & {\href{T4.3.1}{1625120}} & {\href{T4.3.2}{1704624}} & {\href{T4.3.3}{1677472}} & {\href{T4.3.4}{1750000}} & {\href{T4.3.5}{14762}} & {\href{T4.3.6}{1625060}} & {\href{T4.3.7}{1677460}}\\

\hline
\end{tabular}
\end{lrbox}

% ----------------------
% Table: tTimeMFormulaSZWStringPlengthO
% ProcessingtimebymethodFormulaSacStringlengthb
% ----------------------
\newsavebox{\tTimeMFormulaSZWStringPlengthO}
\begin{lrbox}{\tTimeMFormulaSZWStringPlengthO}
\begin{tabular}{|c|c|c|c|c|c|c|c|}
\hline
\textbf{Trace length} & \textbf{EWAH64} & \textbf{Roaring} & \textbf{Concise} & \textbf{Raw} & \textbf{BeepBeep} & \textbf{EWAH32} & \textbf{WAH}\\
\hline\hline
{\href{T7.0.0}{250000}} & {\href{T7.0.1}{2551020.5}} & {\href{T7.0.2}{2577319.5}} & {\href{T7.0.3}{2314814.8}} & {\href{T7.0.4}{2500000.0}} & {\href{T7.0.5}{437062.94}} & {\href{T7.0.6}{2577319.5}} & {\href{T7.0.7}{2976190.5}}\\
\hline
{\href{T7.1.0}{500000}} & {\href{T7.1.1}{2551020.5}} & {\href{T7.1.2}{2392344.5}} & {\href{T7.1.3}{2590673.5}} & {\href{T7.1.4}{2857142.8}} & {\href{T7.1.5}{378501.12}} & {\href{T7.1.6}{2808988.8}} & {\href{T7.1.7}{2538071.0}}\\
\hline
{\href{T7.2.0}{750000}} & {\href{T7.2.1}{2717391.2}} & {\href{T7.2.2}{2435065.0}} & {\href{T7.2.3}{2895753.0}} & {\href{T7.2.4}{3177966.0}} & {\href{T7.2.5}{432027.66}} & {\href{T7.2.6}{2808988.8}} & {\href{T7.2.7}{2964427.0}}\\
\hline
{\href{T7.3.0}{1000000}} & {\href{T7.3.1}{2754821.0}} & {\href{T7.3.2}{2415459.0}} & {\href{T7.3.3}{2949852.5}} & {\href{T7.3.4}{3076923.0}} & {\href{T7.3.5}{444247.0}} & {\href{T7.3.6}{2832861.2}} & {\href{T7.3.7}{2695417.8}}\\

\hline
\end{tabular}
\end{lrbox}

% ----------------------
% Table: tMemMFormulaSZWStringPlengthO
% MemoryconsumptionbymethodFormulaSacStringlengthb
% ----------------------
\newsavebox{\tMemMFormulaSZWStringPlengthO}
\begin{lrbox}{\tMemMFormulaSZWStringPlengthO}
\begin{tabular}{|c|c|c|c|c|c|c|c|}
\hline
\textbf{Trace length} & \textbf{EWAH64} & \textbf{Roaring} & \textbf{Concise} & \textbf{Raw} & \textbf{BeepBeep} & \textbf{EWAH32} & \textbf{WAH}\\
\hline\hline
{\href{T8.0.0}{250000}} & {\href{T8.0.1}{437696}} & {\href{T8.0.2}{458976}} & {\href{T8.0.3}{451640}} & {\href{T8.0.4}{468750}} & {\href{T8.0.5}{5025440}} & {\href{T8.0.6}{437584}} & {\href{T8.0.7}{451640}}\\
\hline
{\href{T8.1.0}{500000}} & {\href{T8.1.1}{875168}} & {\href{T8.1.2}{917856}} & {\href{T8.1.3}{903256}} & {\href{T8.1.4}{937500}} & {\href{T8.1.5}{10025445}} & {\href{T8.1.6}{875056}} & {\href{T8.1.7}{903252}}\\
\hline
{\href{T8.2.0}{750000}} & {\href{T8.2.1}{1312640}} & {\href{T8.2.2}{1376712}} & {\href{T8.2.3}{1354864}} & {\href{T8.2.4}{1406250}} & {\href{T8.2.5}{15025460}} & {\href{T8.2.6}{1312584}} & {\href{T8.2.7}{1354864}}\\
\hline
{\href{T8.3.0}{1000000}} & {\href{T8.3.1}{1750112}} & {\href{T8.3.2}{1835568}} & {\href{T8.3.3}{1806492}} & {\href{T8.3.4}{1875000}} & {\href{T8.3.5}{20025445}} & {\href{T8.3.6}{1750056}} & {\href{T8.3.7}{1806504}}\\

\hline
\end{tabular}
\end{lrbox}

% ----------------------
% Table: tThroughputLongestM
% ProcessingtimeforlongesttracebymethodTracelengthbaaaaaaStringlengthb
% ----------------------
\newsavebox{\tThroughputLongestM}
\begin{lrbox}{\tThroughputLongestM}
\begin{tabular}{|c|c|c|c|c|c|c|c|c|c|}
\hline
\textbf{Formula} & \textbf{Formula size} & \textbf{Formula depth} & \textbf{EWAH64} & \textbf{Roaring} & \textbf{Concise} & \textbf{Raw} & \textbf{BeepBeep} & \textbf{EWAH32} & \textbf{WAH}\\
\hline\hline
{\href{T11.10.0}{A1}} & {\href{T11.10.1}{1}} & {\href{T11.10.2}{1}} & {\href{T11.10.3}{2747252.8}} & {\href{T11.10.4}{2865329.5}} & {\href{T11.10.5}{2958580.0}} & {\href{T11.10.6}{3389830.5}} & {\href{T11.10.7}{2881844.5}} & {\href{T11.10.8}{2659574.5}} & {\href{T11.10.9}{2985074.8}}\\
\hline
{\href{T11.11.0}{A2}} & {\href{T11.11.1}{1}} & {\href{T11.11.2}{1}} & {\href{T11.11.3}{2242152.5}} & {\href{T11.11.4}{2793296.0}} & {\href{T11.11.5}{2873563.2}} & {\href{T11.11.6}{3164557.0}} & {\href{T11.11.7}{2169197.5}} & {\href{T11.11.8}{2739726.0}} & {\href{T11.11.9}{2832861.2}}\\
\hline
{\href{T11.12.0}{A3}} & {\href{T11.12.1}{1}} & {\href{T11.12.2}{1}} & {\href{T11.12.3}{2680965.2}} & {\href{T11.12.4}{3134796.2}} & {\href{T11.12.5}{2881844.5}} & {\href{T11.12.6}{3322259.2}} & {\href{T11.12.7}{2192982.5}} & {\href{T11.12.8}{2739726.0}} & {\href{T11.12.9}{3058104.0}}\\
\hline
{\href{T11.13.0}{A4}} & {\href{T11.13.1}{1}} & {\href{T11.13.2}{1}} & {\href{T11.13.3}{2762431.0}} & {\href{T11.13.4}{2958580.0}} & {\href{T11.13.5}{2873563.2}} & {\href{T11.13.6}{3472222.2}} & {\href{T11.13.7}{2747252.8}} & {\href{T11.13.8}{2617801.0}} & {\href{T11.13.9}{3003003.0}}\\
\hline
{\href{T11.14.0}{A5}} & {\href{T11.14.1}{1}} & {\href{T11.14.2}{1}} & {\href{T11.14.3}{2645502.8}} & {\href{T11.14.4}{2824858.8}} & {\href{T11.14.5}{2881844.5}} & {\href{T11.14.6}{3205128.2}} & {\href{T11.14.7}{2770083.0}} & {\href{T11.14.8}{2770083.0}} & {\href{T11.14.9}{2832861.2}}\\
\hline
{\href{T11.16.0}{A6}} & {\href{T11.16.1}{1}} & {\href{T11.16.2}{1}} & {\href{T11.16.3}{2659574.5}} & {\href{T11.16.4}{2832861.2}} & {\href{T11.16.5}{2857142.8}} & {\href{T11.16.6}{3164557.0}} & {\href{T11.16.7}{2865329.5}} & {\href{T11.16.8}{2849002.8}} & {\href{T11.16.9}{2890173.5}}\\
\hline
\end{tabular}
\end{lrbox}

% ----------------------
% Table: tThroughputLongestRelTracePlengthOZZZZZZStringPlengthO
% RelativeprocessingtimetoBeepBeepTracelengthbaaaaaaStringlengthb
% ----------------------
\newsavebox{\tThroughputLongestRelTracePlengthOZZZZZZStringPlengthO}
\begin{lrbox}{\tThroughputLongestRelTracePlengthOZZZZZZStringPlengthO}
\begin{tabular}{|c|c|c|c|c|c|}
\hline
\textbf{Concise} & \textbf{EWAH32} & \textbf{EWAH64} & \textbf{Raw} & \textbf{Roaring} & \textbf{WAH}\\
\hline\hline
{\href{T13.16.0}{0.99714285}} & {\href{T13.16.1}{0.994302}} & {\href{T13.16.2}{0.9281915}} & {\href{T13.16.3}{1.1044304}} & {\href{T13.16.4}{0.98866856}} & {\href{T13.16.5}{1.0086706}}\\
\hline
 \multirow{5}{*}{\href{T13.13.0}{1.0266272}} & {\href{T13.10.1}{0.9228723}} & {\href{T13.10.2}{0.95329666}} & {\href{T13.10.3}{1.1762711}} & {\href{T13.10.4}{0.9942693}} & {\href{T13.10.5}{1.0358208}}\\
\cline{2-6}
 & {\href{T13.13.1}{0.95287955}} & {\href{T13.13.2}{1.0055249}} & {\href{T13.46.3}{1.2315435}} & {\href{T13.46.4}{1.0110192}} & {\href{T13.46.5}{1.0027323}}\\
\cline{4-6}
 &  &  & {\href{T13.13.3}{1.263889}} & {\href{T13.13.4}{1.0769231}} & {\href{T13.13.5}{1.093093}}\\
\cline{2-6}
 & {\href{T13.14.1}{1.0}} & {\href{T13.14.2}{0.9550265}} & {\href{T13.14.3}{1.1570513}} & {\href{T13.14.4}{1.0197741}} & {\href{T13.14.5}{1.022663}}\\
\cline{3-6}
 &  & {\href{T13.11.2}{1.0336323}} & {\href{T13.11.3}{1.4588608}} & {\href{T13.11.4}{1.2877094}} & {\href{T13.11.5}{1.305949}}\\
\hline
{\href{T13.5.0}{1.7002966}} & {\href{T13.5.1}{1.5118734}} & {\href{T13.5.2}{1.6005585}} & {\href{T13.5.3}{1.8973509}} & {\href{T13.5.4}{1.7155688}} & {\href{T13.5.5}{1.6371428}}\\
\hline
\end{tabular}
\end{lrbox}

% ----------------------
% Table: RelativememoryconsumptiontoBeepBeepboxplot
% RelativememoryconsumptiontoBeepBeepboxplot
% ----------------------
\newsavebox{\RelativememoryconsumptiontoBeepBeepboxplot}
\begin{lrbox}{\RelativememoryconsumptiontoBeepBeepboxplot}
\begin{tabular}{|c|c|c|c|c|c|c|}
\hline
\textbf{x} & \textbf{Min} & \textbf{Q1} & \textbf{Q2} & \textbf{Q3} & \textbf{Max} & \textbf{Label}\\
\hline\hline
{\href{T14.0.0}{0}} & {\href{T14.0.1}{0.99714285}} & {\href{T14.0.2}{1.7002966}} & {\href{T14.0.3}{2.3555555}} & {\href{T14.0.4}{3.0424628}} & {\href{T14.0.5}{8.204445}} & {\href{T14.0.6}{Concise}}\\
\hline
{\href{T14.1.0}{1}} & {\href{T14.1.1}{0.9228723}} & {\href{T14.1.2}{1.5118734}} & {\href{T14.1.3}{2.1324325}} & {\href{T14.1.4}{2.6944444}} & {\href{T14.1.5}{8.982968}} & {\href{T14.1.6}{EWAH32}}\\
\hline
{\href{T14.2.0}{2}} & {\href{T14.2.1}{0.9281915}} & {\href{T14.2.2}{1.5165288}} & {\href{T14.2.3}{2.099825}} & {\href{T14.2.4}{2.6603377}} & {\href{T14.2.5}{8.168142}} & {\href{T14.2.6}{EWAH64}}\\
\hline
{\href{T14.3.0}{3}} & {\href{T14.3.1}{1.1044304}} & {\href{T14.3.2}{1.8947369}} & {\href{T14.3.3}{2.8079472}} & {\href{T14.3.4}{3.433775}} & {\href{T14.3.5}{10.795321}} & {\href{T14.3.6}{Raw}}\\
\hline
{\href{T14.4.0}{4}} & {\href{T14.4.1}{0.98866856}} & {\href{T14.4.2}{1.7021277}} & {\href{T14.4.3}{2.313783}} & {\href{T14.4.4}{2.9531612}} & {\href{T14.4.5}{8.334085}} & {\href{T14.4.6}{Roaring}}\\
\hline
{\href{T14.5.0}{5}} & {\href{T14.5.1}{0.47208124}} & {\href{T14.5.2}{1.6214952}} & {\href{T14.5.3}{2.2162921}} & {\href{T14.5.4}{2.9326317}} & {\href{T14.5.5}{9.323233}} & {\href{T14.5.6}{WAH}}\\

\hline
\end{tabular}
\end{lrbox}

% ----------------------
% Table: tMemLongestM
% MemoryconsumptionforlongesttracebymethodTracelengthbaaaaaaStringlengthb
% ----------------------
\newsavebox{\tMemLongestM}
\begin{lrbox}{\tMemLongestM}
\begin{tabular}{|c|c|c|c|c|c|c|c|c|c|}
\hline
\textbf{Formula} & \textbf{Formula size} & \textbf{Formula depth} & \textbf{EWAH64} & \textbf{Roaring} & \textbf{Concise} & \textbf{Raw} & \textbf{BeepBeep} & \textbf{EWAH32} & \textbf{WAH}\\
\hline\hline
{\href{T15.10.0}{A1}} & {\href{T15.10.1}{1}} & {\href{T15.10.2}{1}} & {\href{T15.10.3}{1250080}} & {\href{T15.10.4}{1311120}} & {\href{T15.10.5}{1290356}} & {\href{T15.10.6}{1250000}} & {\href{T15.10.7}{3417}} & {\href{T15.10.8}{1250040}} & {\href{T15.10.9}{1290352}}\\
\hline
{\href{T15.11.0}{A2}} & {\href{T15.11.1}{1}} & {\href{T15.11.2}{1}} & {\href{T15.11.3}{1500096}} & {\href{T15.11.4}{1573344}} & {\href{T15.11.5}{1548420}} & {\href{T15.11.6}{1500000}} & {\href{T15.11.7}{4556}} & {\href{T15.11.8}{1500048}} & {\href{T15.11.9}{1548412}}\\
\hline
{\href{T15.12.0}{A3}} & {\href{T15.12.1}{1}} & {\href{T15.12.2}{1}} & {\href{T15.12.3}{1500096}} & {\href{T15.12.4}{1573344}} & {\href{T15.12.5}{1548420}} & {\href{T15.12.6}{1500000}} & {\href{T15.12.7}{4556}} & {\href{T15.12.8}{1500048}} & {\href{T15.12.9}{1548432}}\\
\hline
{\href{T15.13.0}{A4}} & {\href{T15.13.1}{1}} & {\href{T15.13.2}{1}} & {\href{T15.13.3}{1250080}} & {\href{T15.13.4}{1311120}} & {\href{T15.13.5}{1290348}} & {\href{T15.13.6}{1250000}} & {\href{T15.13.7}{3364}} & {\href{T15.13.8}{1250040}} & {\href{T15.13.9}{1290340}}\\
\hline
{\href{T15.14.0}{A5}} & {\href{T15.14.1}{1}} & {\href{T15.14.2}{1}} & {\href{T15.14.3}{1250080}} & {\href{T15.14.4}{1311120}} & {\href{T15.14.5}{1290356}} & {\href{T15.14.6}{1250000}} & {\href{T15.14.7}{3364}} & {\href{T15.14.8}{1250040}} & {\href{T15.14.9}{1290360}}\\
\hline
{\href{T15.16.0}{A6}} & {\href{T15.16.1}{1}} & {\href{T15.16.2}{1}} & {\href{T15.16.3}{1250080}} & {\href{T15.16.4}{1311120}} & {\href{T15.16.5}{1290352}} & {\href{T15.16.6}{1250000}} & {\href{T15.16.7}{3364}} & {\href{T15.16.8}{1250040}} & {\href{T15.16.9}{1290344}}\\
\hline
\end{tabular}
\end{lrbox}

% ----------------------
% Table: tMemLongestRelTracePlengthOZZZZZZStringPlengthO
% RelativememoryconsumptiontoBeepBeepTracelengthbaaaaaaStringlengthb
% ----------------------
\newsavebox{\tMemLongestRelTracePlengthOZZZZZZStringPlengthO}
\begin{lrbox}{\tMemLongestRelTracePlengthOZZZZZZStringPlengthO}
\begin{tabular}{|c|c|c|c|c|c|}
\hline
\textbf{Concise} & \textbf{EWAH32} & \textbf{EWAH64} & \textbf{Raw} & \textbf{Roaring} & \textbf{WAH}\\
\hline\hline
{\href{T17.25.0}{0.07736354}} & {\href{T17.25.1}{0.07494636}} & {\href{T17.25.2}{0.07494896}} & {\href{T17.25.3}{0.08118907}} & {\href{T17.25.4}{0.0851589}} & {\href{T17.25.5}{0.07736335}}\\
\hline
{\href{T17.15.0}{0.090209834}} & {\href{T17.15.1}{0.08739162}} & {\href{T17.15.2}{0.08739442}} & {\href{T17.15.3}{0.09363088}} & {\href{T17.15.4}{0.09166179}} & {\href{T17.15.5}{0.09021044}}\\
\hline
{\href{T17.7.0}{0.30916107}} & {\href{T17.7.1}{0.29950225}} & {\href{T17.7.2}{0.29951423}} & {\href{T17.7.3}{0.32444954}} & {\href{T17.7.4}{0.34031382}} & {\href{T17.7.5}{0.30916107}}\\
\hline
{\href{T17.24.0}{88.293106}} & {\href{T17.24.1}{85.53521}} & {\href{T17.24.2}{85.53869}} & {\href{T17.24.3}{108.857925}} & {\href{T17.24.4}{106.02588}} & {\href{T17.24.5}{88.29336}}\\
\hline
{\href{T17.49.0}{89.33991}} & {\href{T17.49.1}{86.549484}} & {\href{T17.49.2}{86.5526}} & {\href{T17.49.3}{91.95551}} & {\href{T17.49.4}{90.77883}} & {\href{T17.49.5}{89.33939}}\\
\hline
{\href{T17.17.0}{113.63447}} & {\href{T17.17.1}{110.084}} & {\href{T17.17.2}{110.088066}} & {\href{T17.17.3}{118.54762}} & {\href{T17.17.4}{115.473785}} & {\href{T17.17.5}{113.63365}}\\
\hline
{\href{T17.29.0}{121.298325}} & {\href{T17.29.1}{117.50916}} & {\href{T17.29.2}{117.513466}} & {\href{T17.29.3}{125.89807}} & {\href{T17.29.4}{123.25173}} & {\href{T17.29.5}{121.298325}}\\
\hline
\end{tabular}
\end{lrbox}

% ----------------------
% Table: Tablebi
% Tablebi
% ----------------------
\newsavebox{\Tablebi}
\begin{lrbox}{\Tablebi}
\begin{tabular}{|c|c|c|c|c|c|c|}
\hline
\textbf{x} & \textbf{Min} & \textbf{Q1} & \textbf{Q2} & \textbf{Q3} & \textbf{Max} & \textbf{Label}\\
\hline\hline
{\href{T18.0.0}{0}} & {\href{T18.0.1}{0.07736354}} & {\href{T18.0.2}{126.02874}} & {\href{T18.0.3}{159.51399}} & {\href{T18.0.4}{236.3929}} & {\href{T18.0.5}{383.57788}} & {\href{T18.0.6}{Concise}}\\
\hline
{\href{T18.1.0}{1}} & {\href{T18.1.1}{0.07494636}} & {\href{T18.1.2}{122.09181}} & {\href{T18.1.3}{154.53104}} & {\href{T18.1.4}{229.00958}} & {\href{T18.1.5}{371.59335}} & {\href{T18.1.6}{EWAH32}}\\
\hline
{\href{T18.2.0}{2}} & {\href{T18.2.1}{0.07494896}} & {\href{T18.2.2}{122.096275}} & {\href{T18.2.3}{154.53598}} & {\href{T18.2.4}{229.0169}} & {\href{T18.2.5}{371.60522}} & {\href{T18.2.6}{EWAH64}}\\
\hline
{\href{T18.3.0}{3}} & {\href{T18.3.1}{0.08118907}} & {\href{T18.3.2}{130.05592}} & {\href{T18.3.3}{165.56291}} & {\href{T18.3.4}{229.00226}} & {\href{T18.3.5}{371.58145}} & {\href{T18.3.6}{Raw}}\\
\hline
{\href{T18.4.0}{4}} & {\href{T18.4.1}{0.0851589}} & {\href{T18.4.2}{128.05748}} & {\href{T18.4.3}{171.8214}} & {\href{T18.4.4}{240.19955}} & {\href{T18.4.5}{389.7503}} & {\href{T18.4.6}{Roaring}}\\
\hline
{\href{T18.5.0}{5}} & {\href{T18.5.1}{0.07736335}} & {\href{T18.5.2}{126.02818}} & {\href{T18.5.3}{159.51434}} & {\href{T18.5.4}{236.39516}} & {\href{T18.5.5}{383.57907}} & {\href{T18.5.6}{WAH}}\\

\hline
\end{tabular}
\end{lrbox}

% ----------------------
% Table: tThroughputLongestDWO
% ProcessingtimebyformulaFormulaDcbTracelengthbaaaaaaStringlengthb
% ----------------------
\newsavebox{\tThroughputLongestDWO}
\begin{lrbox}{\tThroughputLongestDWO}
\begin{tabular}{|c|c|c|c|c|c|c|c|}
\hline
\textbf{Trace length} & \textbf{EWAH64} & \textbf{Roaring} & \textbf{Concise} & \textbf{Raw} & \textbf{BeepBeep} & \textbf{EWAH32} & \textbf{WAH}\\
\hline\hline
{\href{T20.0.0}{1000000}} & {\href{T20.0.1}{2096436.0}} & {\href{T20.0.2}{2304147.5}} & {\href{T20.0.3}{2364066.2}} & {\href{T20.0.4}{2369668.2}} & {\href{T20.0.5}{634920.6}} & {\href{T20.0.6}{2145922.8}} & {\href{T20.0.7}{2267573.8}}\\

\hline
\end{tabular}
\end{lrbox}

% ----------------------
% Table: tThroughputLongestDWZ
% ProcessingtimebyformulaFormulaDcaTracelengthbaaaaaaStringlengthb
% ----------------------
\newsavebox{\tThroughputLongestDWZ}
\begin{lrbox}{\tThroughputLongestDWZ}
\begin{tabular}{|c|c|c|c|c|c|c|c|}
\hline
\textbf{Trace length} & \textbf{EWAH64} & \textbf{Roaring} & \textbf{Concise} & \textbf{Raw} & \textbf{BeepBeep} & \textbf{EWAH32} & \textbf{WAH}\\
\hline\hline
{\href{T22.0.0}{1000000}} & {\href{T22.0.1}{2074688.8}} & {\href{T22.0.2}{2415459.0}} & {\href{T22.0.3}{2024291.5}} & {\href{T22.0.4}{2785515.2}} & {\href{T22.0.5}{1162790.8}} & {\href{T22.0.6}{1385041.5}} & {\href{T22.0.7}{2325581.5}}\\

\hline
\end{tabular}
\end{lrbox}

% ----------------------
% Table: tThroughputLongestDWR
% ProcessingtimebyformulaFormulaDcdTracelengthbaaaaaaStringlengthb
% ----------------------
\newsavebox{\tThroughputLongestDWR}
\begin{lrbox}{\tThroughputLongestDWR}
\begin{tabular}{|c|c|c|c|c|c|c|c|}
\hline
\textbf{Trace length} & \textbf{EWAH64} & \textbf{Roaring} & \textbf{Concise} & \textbf{Raw} & \textbf{BeepBeep} & \textbf{EWAH32} & \textbf{WAH}\\
\hline\hline
{\href{T24.0.0}{1000000}} & {\href{T24.0.1}{2309468.8}} & {\href{T24.0.2}{2415459.0}} & {\href{T24.0.3}{2597402.5}} & {\href{T24.0.4}{2923976.5}} & {\href{T24.0.5}{1169590.6}} & {\href{T24.0.6}{2358490.5}} & {\href{T24.0.7}{2577319.5}}\\

\hline
\end{tabular}
\end{lrbox}

% ----------------------
% Table: tThroughputLongestDWW
% ProcessingtimebyformulaFormulaDccTracelengthbaaaaaaStringlengthb
% ----------------------
\newsavebox{\tThroughputLongestDWW}
\begin{lrbox}{\tThroughputLongestDWW}
\begin{tabular}{|c|c|c|c|c|c|c|c|}
\hline
\textbf{Trace length} & \textbf{EWAH64} & \textbf{Roaring} & \textbf{Concise} & \textbf{Raw} & \textbf{BeepBeep} & \textbf{EWAH32} & \textbf{WAH}\\
\hline\hline
{\href{T26.0.0}{1000000}} & {\href{T26.0.1}{2341920.2}} & {\href{T26.0.2}{2439024.5}} & {\href{T26.0.3}{2444987.8}} & {\href{T26.0.4}{2958580.0}} & {\href{T26.0.5}{917431.2}} & {\href{T26.0.6}{2212389.5}} & {\href{T26.0.7}{2386634.8}}\\

\hline
\end{tabular}
\end{lrbox}

% ----------------------
% Table: tThroughputLongestDWV
% ProcessingtimebyformulaFormulaDcfTracelengthbaaaaaaStringlengthb
% ----------------------
\newsavebox{\tThroughputLongestDWV}
\begin{lrbox}{\tThroughputLongestDWV}
\begin{tabular}{|c|c|c|c|c|c|c|c|}
\hline
\textbf{Trace length} & \textbf{EWAH64} & \textbf{Roaring} & \textbf{Concise} & \textbf{Raw} & \textbf{BeepBeep} & \textbf{EWAH32} & \textbf{WAH}\\
\hline\hline
{\href{T28.0.0}{1000000}} & {\href{T28.0.1}{1792114.8}} & {\href{T28.0.2}{1930501.9}} & {\href{T28.0.3}{2079002.1}} & {\href{T28.0.4}{2512562.8}} & {\href{T28.0.5}{843881.9}} & {\href{T28.0.6}{1941747.6}} & {\href{T28.0.7}{2173913.0}}\\

\hline
\end{tabular}
\end{lrbox}

% ----------------------
% Table: tThroughputLongestDWF
% ProcessingtimebyformulaFormulaDceTracelengthbaaaaaaStringlengthb
% ----------------------
\newsavebox{\tThroughputLongestDWF}
\begin{lrbox}{\tThroughputLongestDWF}
\begin{tabular}{|c|c|c|c|c|c|c|c|}
\hline
\textbf{Trace length} & \textbf{EWAH64} & \textbf{Roaring} & \textbf{Concise} & \textbf{Raw} & \textbf{BeepBeep} & \textbf{EWAH32} & \textbf{WAH}\\
\hline\hline
{\href{T30.0.0}{1000000}} & {\href{T30.0.1}{2793296.0}} & {\href{T30.0.2}{2994012.0}} & {\href{T30.0.3}{2967359.0}} & {\href{T30.0.4}{3311258.2}} & {\href{T30.0.5}{1745200.8}} & {\href{T30.0.6}{2638522.5}} & {\href{T30.0.7}{2857142.8}}\\

\hline
\end{tabular}
\end{lrbox}

% ----------------------
% Table: tThroughputLongestDWS
% ProcessingtimebyformulaFormulaDchTracelengthbaaaaaaStringlengthb
% ----------------------
\newsavebox{\tThroughputLongestDWS}
\begin{lrbox}{\tThroughputLongestDWS}
\begin{tabular}{|c|c|c|c|c|c|c|c|}
\hline
\textbf{Trace length} & \textbf{EWAH64} & \textbf{Roaring} & \textbf{Concise} & \textbf{Raw} & \textbf{BeepBeep} & \textbf{EWAH32} & \textbf{WAH}\\
\hline\hline
{\href{T32.0.0}{1000000}} & {\href{T32.0.1}{2109704.8}} & {\href{T32.0.2}{2341920.2}} & {\href{T32.0.3}{2518891.8}} & {\href{T32.0.4}{2865329.5}} & {\href{T32.0.5}{793021.44}} & {\href{T32.0.6}{2136752.2}} & {\href{T32.0.7}{2358490.5}}\\

\hline
\end{tabular}
\end{lrbox}

% ----------------------
% Table: tThroughputLongestDWX
% ProcessingtimebyformulaFormulaDcgTracelengthbaaaaaaStringlengthb
% ----------------------
\newsavebox{\tThroughputLongestDWX}
\begin{lrbox}{\tThroughputLongestDWX}
\begin{tabular}{|c|c|c|c|c|c|c|c|}
\hline
\textbf{Trace length} & \textbf{EWAH64} & \textbf{Roaring} & \textbf{Concise} & \textbf{Raw} & \textbf{BeepBeep} & \textbf{EWAH32} & \textbf{WAH}\\
\hline\hline
{\href{T34.0.0}{1000000}} & {\href{T34.0.1}{2604166.8}} & {\href{T34.0.2}{2932551.2}} & {\href{T34.0.3}{3058104.0}} & {\href{T34.0.4}{3389830.5}} & {\href{T34.0.5}{1267427.1}} & {\href{T34.0.6}{2702702.8}} & {\href{T34.0.7}{2808988.8}}\\

\hline
\end{tabular}
\end{lrbox}

% ----------------------
% Table: tThroughputLongestDWN
% ProcessingtimebyformulaFormulaDcjTracelengthbaaaaaaStringlengthb
% ----------------------
\newsavebox{\tThroughputLongestDWN}
\begin{lrbox}{\tThroughputLongestDWN}
\begin{tabular}{|c|c|c|c|c|c|c|c|}
\hline
\textbf{Trace length} & \textbf{EWAH64} & \textbf{Roaring} & \textbf{Concise} & \textbf{Raw} & \textbf{BeepBeep} & \textbf{EWAH32} & \textbf{WAH}\\
\hline\hline
{\href{T36.0.0}{1000000}} & {\href{T36.0.1}{1754386.0}} & {\href{T36.0.2}{1945525.2}} & {\href{T36.0.3}{1949317.8}} & {\href{T36.0.4}{2386634.8}} & {\href{T36.0.5}{717875.06}} & {\href{T36.0.6}{1834862.4}} & {\href{T36.0.7}{2105263.2}}\\

\hline
\end{tabular}
\end{lrbox}

% ----------------------
% Table: tThroughputLongestDWH
% ProcessingtimebyformulaFormulaDciTracelengthbaaaaaaStringlengthb
% ----------------------
\newsavebox{\tThroughputLongestDWH}
\begin{lrbox}{\tThroughputLongestDWH}
\begin{tabular}{|c|c|c|c|c|c|c|c|}
\hline
\textbf{Trace length} & \textbf{EWAH64} & \textbf{Roaring} & \textbf{Concise} & \textbf{Raw} & \textbf{BeepBeep} & \textbf{EWAH32} & \textbf{WAH}\\
\hline\hline
{\href{T38.0.0}{1000000}} & {\href{T38.0.1}{1751313.5}} & {\href{T38.0.2}{1976284.6}} & {\href{T38.0.3}{2079002.1}} & {\href{T38.0.4}{2369668.2}} & {\href{T38.0.5}{834028.4}} & {\href{T38.0.6}{1862197.4}} & {\href{T38.0.7}{2012072.4}}\\

\hline
\end{tabular}
\end{lrbox}

% ----------------------
% Table: tThroughputLongestAO
% ProcessingtimebyformulaFormulaAbTracelengthbaaaaaaStringlengthb
% ----------------------
\newsavebox{\tThroughputLongestAO}
\begin{lrbox}{\tThroughputLongestAO}
\begin{tabular}{|c|c|c|c|c|c|c|c|}
\hline
\textbf{Trace length} & \textbf{EWAH64} & \textbf{Roaring} & \textbf{Concise} & \textbf{Raw} & \textbf{BeepBeep} & \textbf{EWAH32} & \textbf{WAH}\\
\hline\hline
{\href{T40.0.0}{1000000}} & {\href{T40.0.1}{2747252.8}} & {\href{T40.0.2}{2865329.5}} & {\href{T40.0.3}{2958580.0}} & {\href{T40.0.4}{3389830.5}} & {\href{T40.0.5}{2881844.5}} & {\href{T40.0.6}{2659574.5}} & {\href{T40.0.7}{2985074.8}}\\

\hline
\end{tabular}
\end{lrbox}

% ----------------------
% Table: tThroughputLongestAW
% ProcessingtimebyformulaFormulaAcTracelengthbaaaaaaStringlengthb
% ----------------------
\newsavebox{\tThroughputLongestAW}
\begin{lrbox}{\tThroughputLongestAW}
\begin{tabular}{|c|c|c|c|c|c|c|c|}
\hline
\textbf{Trace length} & \textbf{EWAH64} & \textbf{Roaring} & \textbf{Concise} & \textbf{Raw} & \textbf{BeepBeep} & \textbf{EWAH32} & \textbf{WAH}\\
\hline\hline
{\href{T42.0.0}{1000000}} & {\href{T42.0.1}{2242152.5}} & {\href{T42.0.2}{2793296.0}} & {\href{T42.0.3}{2873563.2}} & {\href{T42.0.4}{3164557.0}} & {\href{T42.0.5}{2169197.5}} & {\href{T42.0.6}{2739726.0}} & {\href{T42.0.7}{2832861.2}}\\

\hline
\end{tabular}
\end{lrbox}

% ----------------------
% Table: tThroughputLongestAR
% ProcessingtimebyformulaFormulaAdTracelengthbaaaaaaStringlengthb
% ----------------------
\newsavebox{\tThroughputLongestAR}
\begin{lrbox}{\tThroughputLongestAR}
\begin{tabular}{|c|c|c|c|c|c|c|c|}
\hline
\textbf{Trace length} & \textbf{EWAH64} & \textbf{Roaring} & \textbf{Concise} & \textbf{Raw} & \textbf{BeepBeep} & \textbf{EWAH32} & \textbf{WAH}\\
\hline\hline
{\href{T44.0.0}{1000000}} & {\href{T44.0.1}{2680965.2}} & {\href{T44.0.2}{3134796.2}} & {\href{T44.0.3}{2881844.5}} & {\href{T44.0.4}{3322259.2}} & {\href{T44.0.5}{2192982.5}} & {\href{T44.0.6}{2739726.0}} & {\href{T44.0.7}{3058104.0}}\\

\hline
\end{tabular}
\end{lrbox}

% ----------------------
% Table: tThroughputLongestAF
% ProcessingtimebyformulaFormulaAeTracelengthbaaaaaaStringlengthb
% ----------------------
\newsavebox{\tThroughputLongestAF}
\begin{lrbox}{\tThroughputLongestAF}
\begin{tabular}{|c|c|c|c|c|c|c|c|}
\hline
\textbf{Trace length} & \textbf{EWAH64} & \textbf{Roaring} & \textbf{Concise} & \textbf{Raw} & \textbf{BeepBeep} & \textbf{EWAH32} & \textbf{WAH}\\
\hline\hline
{\href{T46.0.0}{1000000}} & {\href{T46.0.1}{2762431.0}} & {\href{T46.0.2}{2958580.0}} & {\href{T46.0.3}{2873563.2}} & {\href{T46.0.4}{3472222.2}} & {\href{T46.0.5}{2747252.8}} & {\href{T46.0.6}{2617801.0}} & {\href{T46.0.7}{3003003.0}}\\

\hline
\end{tabular}
\end{lrbox}

% ----------------------
% Table: tThroughputLongestAV
% ProcessingtimebyformulaFormulaAfTracelengthbaaaaaaStringlengthb
% ----------------------
\newsavebox{\tThroughputLongestAV}
\begin{lrbox}{\tThroughputLongestAV}
\begin{tabular}{|c|c|c|c|c|c|c|c|}
\hline
\textbf{Trace length} & \textbf{EWAH64} & \textbf{Roaring} & \textbf{Concise} & \textbf{Raw} & \textbf{BeepBeep} & \textbf{EWAH32} & \textbf{WAH}\\
\hline\hline
{\href{T48.0.0}{1000000}} & {\href{T48.0.1}{2645502.8}} & {\href{T48.0.2}{2824858.8}} & {\href{T48.0.3}{2881844.5}} & {\href{T48.0.4}{3205128.2}} & {\href{T48.0.5}{2770083.0}} & {\href{T48.0.6}{2770083.0}} & {\href{T48.0.7}{2832861.2}}\\

\hline
\end{tabular}
\end{lrbox}

% ----------------------
% Table: tThroughputLongestSZW
% ProcessingtimebyformulaFormulaSacTracelengthbaaaaaaStringlengthb
% ----------------------
\newsavebox{\tThroughputLongestSZW}
\begin{lrbox}{\tThroughputLongestSZW}
\begin{tabular}{|c|c|c|c|c|c|c|c|}
\hline
\textbf{Trace length} & \textbf{EWAH64} & \textbf{Roaring} & \textbf{Concise} & \textbf{Raw} & \textbf{BeepBeep} & \textbf{EWAH32} & \textbf{WAH}\\
\hline\hline
{\href{T50.0.0}{1000000}} & {\href{T50.0.1}{2754821.0}} & {\href{T50.0.2}{2415459.0}} & {\href{T50.0.3}{2949852.5}} & {\href{T50.0.4}{3076923.0}} & {\href{T50.0.5}{444247.0}} & {\href{T50.0.6}{2832861.2}} & {\href{T50.0.7}{2695417.8}}\\

\hline
\end{tabular}
\end{lrbox}

% ----------------------
% Table: tThroughputLongestAX
% ProcessingtimebyformulaFormulaAgTracelengthbaaaaaaStringlengthb
% ----------------------
\newsavebox{\tThroughputLongestAX}
\begin{lrbox}{\tThroughputLongestAX}
\begin{tabular}{|c|c|c|c|c|c|c|c|}
\hline
\textbf{Trace length} & \textbf{EWAH64} & \textbf{Roaring} & \textbf{Concise} & \textbf{Raw} & \textbf{BeepBeep} & \textbf{EWAH32} & \textbf{WAH}\\
\hline\hline
{\href{T52.0.0}{1000000}} & {\href{T52.0.1}{2659574.5}} & {\href{T52.0.2}{2832861.2}} & {\href{T52.0.3}{2857142.8}} & {\href{T52.0.4}{3164557.0}} & {\href{T52.0.5}{2865329.5}} & {\href{T52.0.6}{2849002.8}} & {\href{T52.0.7}{2890173.5}}\\

\hline
\end{tabular}
\end{lrbox}

% ----------------------
% Table: tThroughputLongestSZO
% ProcessingtimebyformulaFormulaSabTracelengthbaaaaaaStringlengthb
% ----------------------
\newsavebox{\tThroughputLongestSZO}
\begin{lrbox}{\tThroughputLongestSZO}
\begin{tabular}{|c|c|c|c|c|c|c|c|}
\hline
\textbf{Trace length} & \textbf{EWAH64} & \textbf{Roaring} & \textbf{Concise} & \textbf{Raw} & \textbf{BeepBeep} & \textbf{EWAH32} & \textbf{WAH}\\
\hline\hline
{\href{T54.0.0}{1000000}} & {\href{T54.0.1}{2617801.0}} & {\href{T54.0.2}{2680965.2}} & {\href{T54.0.3}{2808988.8}} & {\href{T54.0.4}{3412969.2}} & {\href{T54.0.5}{693000.7}} & {\href{T54.0.6}{2777777.8}} & {\href{T54.0.7}{2832861.2}}\\

\hline
\end{tabular}
\end{lrbox}

% ----------------------
% Table: tThroughputLongestAS
% ProcessingtimebyformulaFormulaAhTracelengthbaaaaaaStringlengthb
% ----------------------
\newsavebox{\tThroughputLongestAS}
\begin{lrbox}{\tThroughputLongestAS}
\begin{tabular}{|c|c|c|c|c|c|c|c|}
\hline
\textbf{Trace length} & \textbf{EWAH64} & \textbf{Roaring} & \textbf{Concise} & \textbf{Raw} & \textbf{BeepBeep} & \textbf{EWAH32} & \textbf{WAH}\\
\hline\hline
{\href{T56.0.0}{1000000}} & {\href{T56.0.1}{2100840.2}} & {\href{T56.0.2}{2369668.2}} & {\href{T56.0.3}{2481389.5}} & {\href{T56.0.4}{2739726.0}} & {\href{T56.0.5}{2141327.5}} & {\href{T56.0.6}{2267573.8}} & {\href{T56.0.7}{2487562.2}}\\

\hline
\end{tabular}
\end{lrbox}

% ----------------------
% Table: tThroughputLongestSZF
% ProcessingtimebyformulaFormulaSaeTracelengthbaaaaaaStringlengthb
% ----------------------
\newsavebox{\tThroughputLongestSZF}
\begin{lrbox}{\tThroughputLongestSZF}
\begin{tabular}{|c|c|c|c|c|c|c|c|}
\hline
\textbf{Trace length} & \textbf{EWAH64} & \textbf{Roaring} & \textbf{Concise} & \textbf{Raw} & \textbf{BeepBeep} & \textbf{EWAH32} & \textbf{WAH}\\
\hline\hline
{\href{T58.0.0}{1000000}} & {\href{T58.0.1}{1785714.2}} & {\href{T58.0.2}{1926782.2}} & {\href{T58.0.3}{2118644.0}} & {\href{T58.0.4}{2518891.8}} & {\href{T58.0.5}{1420454.5}} & {\href{T58.0.6}{1834862.4}} & {\href{T58.0.7}{1984127.0}}\\

\hline
\end{tabular}
\end{lrbox}

% ----------------------
% Table: tThroughputLongestSZR
% ProcessingtimebyformulaFormulaSadTracelengthbaaaaaaStringlengthb
% ----------------------
\newsavebox{\tThroughputLongestSZR}
\begin{lrbox}{\tThroughputLongestSZR}
\begin{tabular}{|c|c|c|c|c|c|c|c|}
\hline
\textbf{Trace length} & \textbf{EWAH64} & \textbf{Roaring} & \textbf{Concise} & \textbf{Raw} & \textbf{BeepBeep} & \textbf{EWAH32} & \textbf{WAH}\\
\hline\hline
{\href{T60.0.0}{1000000}} & {\href{T60.0.1}{2247191.0}} & {\href{T60.0.2}{2481389.5}} & {\href{T60.0.3}{2475247.5}} & {\href{T60.0.4}{2949852.5}} & {\href{T60.0.5}{1194743.1}} & {\href{T60.0.6}{2267573.8}} & {\href{T60.0.7}{564015.8}}\\

\hline
\end{tabular}
\end{lrbox}

% ----------------------
% Table: tThroughputLongestSZX
% ProcessingtimebyformulaFormulaSagTracelengthbaaaaaaStringlengthb
% ----------------------
\newsavebox{\tThroughputLongestSZX}
\begin{lrbox}{\tThroughputLongestSZX}
\begin{tabular}{|c|c|c|c|c|c|c|c|}
\hline
\textbf{Trace length} & \textbf{EWAH64} & \textbf{Roaring} & \textbf{Concise} & \textbf{Raw} & \textbf{BeepBeep} & \textbf{EWAH32} & \textbf{WAH}\\
\hline\hline
{\href{T62.0.0}{1000000}} & {\href{T62.0.1}{1610306.0}} & {\href{T62.0.2}{1602564.1}} & {\href{T62.0.3}{1949317.8}} & {\href{T62.0.4}{1795332.1}} & {\href{T62.0.5}{403551.25}} & {\href{T62.0.6}{1754386.0}} & {\href{T62.0.7}{1633986.9}}\\

\hline
\end{tabular}
\end{lrbox}

% ----------------------
% Table: tThroughputLongestSZV
% ProcessingtimebyformulaFormulaSafTracelengthbaaaaaaStringlengthb
% ----------------------
\newsavebox{\tThroughputLongestSZV}
\begin{lrbox}{\tThroughputLongestSZV}
\begin{tabular}{|c|c|c|c|c|c|c|c|}
\hline
\textbf{Trace length} & \textbf{EWAH64} & \textbf{Roaring} & \textbf{Concise} & \textbf{Raw} & \textbf{BeepBeep} & \textbf{EWAH32} & \textbf{WAH}\\
\hline\hline
{\href{T64.0.0}{1000000}} & {\href{T64.0.1}{2212389.5}} & {\href{T64.0.2}{2257336.2}} & {\href{T64.0.3}{2222222.2}} & {\href{T64.0.4}{2923976.5}} & {\href{T64.0.5}{270855.9}} & {\href{T64.0.6}{2433090.0}} & {\href{T64.0.7}{2525252.5}}\\

\hline
\end{tabular}
\end{lrbox}

% ----------------------
% Table: tThroughputLongestSZH
% ProcessingtimebyformulaFormulaSaiTracelengthbaaaaaaStringlengthb
% ----------------------
\newsavebox{\tThroughputLongestSZH}
\begin{lrbox}{\tThroughputLongestSZH}
\begin{tabular}{|c|c|c|c|c|c|c|c|}
\hline
\textbf{Trace length} & \textbf{EWAH64} & \textbf{Roaring} & \textbf{Concise} & \textbf{Raw} & \textbf{BeepBeep} & \textbf{EWAH32} & \textbf{WAH}\\
\hline\hline
{\href{T66.0.0}{1000000}} & {\href{T66.0.1}{1418439.8}} & {\href{T66.0.2}{1555210.0}} & {\href{T66.0.3}{1631321.4}} & {\href{T66.0.4}{1945525.2}} & {\href{T66.0.5}{612745.1}} & {\href{T66.0.6}{1543209.9}} & {\href{T66.0.7}{1661129.6}}\\

\hline
\end{tabular}
\end{lrbox}

% ----------------------
% Table: tThroughputLongestSZS
% ProcessingtimebyformulaFormulaSahTracelengthbaaaaaaStringlengthb
% ----------------------
\newsavebox{\tThroughputLongestSZS}
\begin{lrbox}{\tThroughputLongestSZS}
\begin{tabular}{|c|c|c|c|c|c|c|c|}
\hline
\textbf{Trace length} & \textbf{EWAH64} & \textbf{Roaring} & \textbf{Concise} & \textbf{Raw} & \textbf{BeepBeep} & \textbf{EWAH32} & \textbf{WAH}\\
\hline\hline
{\href{T68.0.0}{1000000}} & {\href{T68.0.1}{2597402.5}} & {\href{T68.0.2}{2518891.8}} & {\href{T68.0.3}{2808988.8}} & {\href{T68.0.4}{3257329.0}} & {\href{T68.0.5}{604960.7}} & {\href{T68.0.6}{2617801.0}} & {\href{T68.0.7}{2881844.5}}\\

\hline
\end{tabular}
\end{lrbox}

% ----------------------
% Table: tThroughputLongestSZN
% ProcessingtimebyformulaFormulaSajTracelengthbaaaaaaStringlengthb
% ----------------------
\newsavebox{\tThroughputLongestSZN}
\begin{lrbox}{\tThroughputLongestSZN}
\begin{tabular}{|c|c|c|c|c|c|c|c|}
\hline
\textbf{Trace length} & \textbf{EWAH64} & \textbf{Roaring} & \textbf{Concise} & \textbf{Raw} & \textbf{BeepBeep} & \textbf{EWAH32} & \textbf{WAH}\\
\hline\hline
{\href{T70.0.0}{1000000}} & {\href{T70.0.1}{1449275.4}} & {\href{T70.0.2}{1492537.4}} & {\href{T70.0.3}{1769911.5}} & {\href{T70.0.4}{1941747.6}} & {\href{T70.0.5}{683994.5}} & {\href{T70.0.6}{1562500.0}} & {\href{T70.0.7}{1739130.4}}\\

\hline
\end{tabular}
\end{lrbox}

% ----------------------
% Table: tThroughputLongestDRW
% ProcessingtimebyformulaFormulaDdcTracelengthbaaaaaaStringlengthb
% ----------------------
\newsavebox{\tThroughputLongestDRW}
\begin{lrbox}{\tThroughputLongestDRW}
\begin{tabular}{|c|c|c|c|c|c|c|c|}
\hline
\textbf{Trace length} & \textbf{EWAH64} & \textbf{Roaring} & \textbf{Concise} & \textbf{Raw} & \textbf{BeepBeep} & \textbf{EWAH32} & \textbf{WAH}\\
\hline\hline
{\href{T72.0.0}{1000000}} & {\href{T72.0.1}{2123142.2}} & {\href{T72.0.2}{2380952.5}} & {\href{T72.0.3}{2320185.5}} & {\href{T72.0.4}{2857142.8}} & {\href{T72.0.5}{980392.2}} & {\href{T72.0.6}{2237136.5}} & {\href{T72.0.7}{2493765.5}}\\

\hline
\end{tabular}
\end{lrbox}

% ----------------------
% Table: tThroughputLongestDRO
% ProcessingtimebyformulaFormulaDdbTracelengthbaaaaaaStringlengthb
% ----------------------
\newsavebox{\tThroughputLongestDRO}
\begin{lrbox}{\tThroughputLongestDRO}
\begin{tabular}{|c|c|c|c|c|c|c|c|}
\hline
\textbf{Trace length} & \textbf{EWAH64} & \textbf{Roaring} & \textbf{Concise} & \textbf{Raw} & \textbf{BeepBeep} & \textbf{EWAH32} & \textbf{WAH}\\
\hline\hline
{\href{T74.0.0}{1000000}} & {\href{T74.0.1}{1757469.2}} & {\href{T74.0.2}{2016129.0}} & {\href{T74.0.3}{2096436.0}} & {\href{T74.0.4}{2257336.2}} & {\href{T74.0.5}{711237.56}} & {\href{T74.0.6}{1893939.4}} & {\href{T74.0.7}{1964636.5}}\\

\hline
\end{tabular}
\end{lrbox}

% ----------------------
% Table: tThroughputLongestDRF
% ProcessingtimebyformulaFormulaDdeTracelengthbaaaaaaStringlengthb
% ----------------------
\newsavebox{\tThroughputLongestDRF}
\begin{lrbox}{\tThroughputLongestDRF}
\begin{tabular}{|c|c|c|c|c|c|c|c|}
\hline
\textbf{Trace length} & \textbf{EWAH64} & \textbf{Roaring} & \textbf{Concise} & \textbf{Raw} & \textbf{BeepBeep} & \textbf{EWAH32} & \textbf{WAH}\\
\hline\hline
{\href{T76.0.0}{1000000}} & {\href{T76.0.1}{1845018.5}} & {\href{T76.0.2}{2020202.0}} & {\href{T76.0.3}{2092050.2}} & {\href{T76.0.4}{2364066.2}} & {\href{T76.0.5}{769822.94}} & {\href{T76.0.6}{1926782.2}} & {\href{T76.0.7}{1976284.6}}\\

\hline
\end{tabular}
\end{lrbox}

% ----------------------
% Table: tThroughputLongestDRR
% ProcessingtimebyformulaFormulaDddTracelengthbaaaaaaStringlengthb
% ----------------------
\newsavebox{\tThroughputLongestDRR}
\begin{lrbox}{\tThroughputLongestDRR}
\begin{tabular}{|c|c|c|c|c|c|c|c|}
\hline
\textbf{Trace length} & \textbf{EWAH64} & \textbf{Roaring} & \textbf{Concise} & \textbf{Raw} & \textbf{BeepBeep} & \textbf{EWAH32} & \textbf{WAH}\\
\hline\hline
{\href{T78.0.0}{1000000}} & {\href{T78.0.1}{1831501.9}} & {\href{T78.0.2}{2020202.0}} & {\href{T78.0.3}{2123142.2}} & {\href{T78.0.4}{2439024.5}} & {\href{T78.0.5}{697836.7}} & {\href{T78.0.6}{1960784.4}} & {\href{T78.0.7}{2169197.5}}\\

\hline
\end{tabular}
\end{lrbox}

% ----------------------
% Table: tThroughputLongestDRH
% ProcessingtimebyformulaFormulaDdiTracelengthbaaaaaaStringlengthb
% ----------------------
\newsavebox{\tThroughputLongestDRH}
\begin{lrbox}{\tThroughputLongestDRH}
\begin{tabular}{|c|c|c|c|c|c|c|c|}
\hline
\textbf{Trace length} & \textbf{EWAH64} & \textbf{Roaring} & \textbf{Concise} & \textbf{Raw} & \textbf{BeepBeep} & \textbf{EWAH32} & \textbf{WAH}\\
\hline\hline
{\href{T80.0.0}{1000000}} & {\href{T80.0.1}{2141327.5}} & {\href{T80.0.2}{2217295.0}} & {\href{T80.0.3}{2336448.5}} & {\href{T80.0.4}{2638522.5}} & {\href{T80.0.5}{845308.56}} & {\href{T80.0.6}{2252252.2}} & {\href{T80.0.7}{2314814.8}}\\

\hline
\end{tabular}
\end{lrbox}

% ----------------------
% Table: tThroughputLongestDRS
% ProcessingtimebyformulaFormulaDdhTracelengthbaaaaaaStringlengthb
% ----------------------
\newsavebox{\tThroughputLongestDRS}
\begin{lrbox}{\tThroughputLongestDRS}
\begin{tabular}{|c|c|c|c|c|c|c|c|}
\hline
\textbf{Trace length} & \textbf{EWAH64} & \textbf{Roaring} & \textbf{Concise} & \textbf{Raw} & \textbf{BeepBeep} & \textbf{EWAH32} & \textbf{WAH}\\
\hline\hline
{\href{T82.0.0}{1000000}} & {\href{T82.0.1}{2785515.2}} & {\href{T82.0.2}{2849002.8}} & {\href{T82.0.3}{2881844.5}} & {\href{T82.0.4}{3311258.2}} & {\href{T82.0.5}{964320.1}} & {\href{T82.0.6}{2724795.8}} & {\href{T82.0.7}{2770083.0}}\\

\hline
\end{tabular}
\end{lrbox}

% ----------------------
% Table: tThroughputLongestDRN
% ProcessingtimebyformulaFormulaDdjTracelengthbaaaaaaStringlengthb
% ----------------------
\newsavebox{\tThroughputLongestDRN}
\begin{lrbox}{\tThroughputLongestDRN}
\begin{tabular}{|c|c|c|c|c|c|c|c|}
\hline
\textbf{Trace length} & \textbf{EWAH64} & \textbf{Roaring} & \textbf{Concise} & \textbf{Raw} & \textbf{BeepBeep} & \textbf{EWAH32} & \textbf{WAH}\\
\hline\hline
{\href{T84.0.0}{1000000}} & {\href{T84.0.1}{2785515.2}} & {\href{T84.0.2}{2724795.8}} & {\href{T84.0.3}{2832861.2}} & {\href{T84.0.4}{3154574.2}} & {\href{T84.0.5}{919963.2}} & {\href{T84.0.6}{2710027.0}} & {\href{T84.0.7}{3039513.8}}\\

\hline
\end{tabular}
\end{lrbox}

% ----------------------
% Table: tThroughputLongestSOO
% ProcessingtimebyformulaFormulaSbbTracelengthbaaaaaaStringlengthb
% ----------------------
\newsavebox{\tThroughputLongestSOO}
\begin{lrbox}{\tThroughputLongestSOO}
\begin{tabular}{|c|c|c|c|c|c|c|c|}
\hline
\textbf{Trace length} & \textbf{EWAH64} & \textbf{Roaring} & \textbf{Concise} & \textbf{Raw} & \textbf{BeepBeep} & \textbf{EWAH32} & \textbf{WAH}\\
\hline\hline
{\href{T86.0.0}{1000000}} & {\href{T86.0.1}{2212389.5}} & {\href{T86.0.2}{2450980.5}} & {\href{T86.0.3}{2341920.2}} & {\href{T86.0.4}{2840909.0}} & {\href{T86.0.5}{1440922.2}} & {\href{T86.0.6}{2242152.5}} & {\href{T86.0.7}{2336448.5}}\\

\hline
\end{tabular}
\end{lrbox}

% ----------------------
% Table: tThroughputLongestSOZ
% ProcessingtimebyformulaFormulaSbaTracelengthbaaaaaaStringlengthb
% ----------------------
\newsavebox{\tThroughputLongestSOZ}
\begin{lrbox}{\tThroughputLongestSOZ}
\begin{tabular}{|c|c|c|c|c|c|c|c|}
\hline
\textbf{Trace length} & \textbf{EWAH64} & \textbf{Roaring} & \textbf{Concise} & \textbf{Raw} & \textbf{BeepBeep} & \textbf{EWAH32} & \textbf{WAH}\\
\hline\hline
{\href{T88.0.0}{1000000}} & {\href{T88.0.1}{2724795.8}} & {\href{T88.0.2}{2247191.0}} & {\href{T88.0.3}{2923976.5}} & {\href{T88.0.4}{3144654.0}} & {\href{T88.0.5}{703729.75}} & {\href{T88.0.6}{2808988.8}} & {\href{T88.0.7}{2770083.0}}\\

\hline
\end{tabular}
\end{lrbox}

% ----------------------
% Table: tThroughputLongestSOR
% ProcessingtimebyformulaFormulaSbdTracelengthbaaaaaaStringlengthb
% ----------------------
\newsavebox{\tThroughputLongestSOR}
\begin{lrbox}{\tThroughputLongestSOR}
\begin{tabular}{|c|c|c|c|c|c|c|c|}
\hline
\textbf{Trace length} & \textbf{EWAH64} & \textbf{Roaring} & \textbf{Concise} & \textbf{Raw} & \textbf{BeepBeep} & \textbf{EWAH32} & \textbf{WAH}\\
\hline\hline
{\href{T90.0.0}{1000000}} & {\href{T90.0.1}{2638522.5}} & {\href{T90.0.2}{2915452.0}} & {\href{T90.0.3}{2849002.8}} & {\href{T90.0.4}{3436426.0}} & {\href{T90.0.5}{921659.0}} & {\href{T90.0.6}{2724795.8}} & {\href{T90.0.7}{3012048.2}}\\

\hline
\end{tabular}
\end{lrbox}

% ----------------------
% Table: tThroughputLongestSOW
% ProcessingtimebyformulaFormulaSbcTracelengthbaaaaaaStringlengthb
% ----------------------
\newsavebox{\tThroughputLongestSOW}
\begin{lrbox}{\tThroughputLongestSOW}
\begin{tabular}{|c|c|c|c|c|c|c|c|}
\hline
\textbf{Trace length} & \textbf{EWAH64} & \textbf{Roaring} & \textbf{Concise} & \textbf{Raw} & \textbf{BeepBeep} & \textbf{EWAH32} & \textbf{WAH}\\
\hline\hline
{\href{T92.0.0}{1000000}} & {\href{T92.0.1}{1858736.0}} & {\href{T92.0.2}{1828153.6}} & {\href{T92.0.3}{2000000.0}} & {\href{T92.0.4}{2427184.5}} & {\href{T92.0.5}{492125.97}} & {\href{T92.0.6}{1901140.6}} & {\href{T92.0.7}{2100840.2}}\\

\hline
\end{tabular}
\end{lrbox}

% ----------------------
% Table: tThroughputLongestDFO
% ProcessingtimebyformulaFormulaDebTracelengthbaaaaaaStringlengthb
% ----------------------
\newsavebox{\tThroughputLongestDFO}
\begin{lrbox}{\tThroughputLongestDFO}
\begin{tabular}{|c|c|c|c|c|c|c|c|}
\hline
\textbf{Trace length} & \textbf{EWAH64} & \textbf{Roaring} & \textbf{Concise} & \textbf{Raw} & \textbf{BeepBeep} & \textbf{EWAH32} & \textbf{WAH}\\
\hline\hline
{\href{T94.0.0}{1000000}} & {\href{T94.0.1}{2192982.5}} & {\href{T94.0.2}{2242152.5}} & {\href{T94.0.3}{2427184.5}} & {\href{T94.0.4}{2544529.2}} & {\href{T94.0.5}{900900.9}} & {\href{T94.0.6}{2252252.2}} & {\href{T94.0.7}{1980198.0}}\\

\hline
\end{tabular}
\end{lrbox}

% ----------------------
% Table: tThroughputLongestDFZ
% ProcessingtimebyformulaFormulaDeaTracelengthbaaaaaaStringlengthb
% ----------------------
\newsavebox{\tThroughputLongestDFZ}
\begin{lrbox}{\tThroughputLongestDFZ}
\begin{tabular}{|c|c|c|c|c|c|c|c|}
\hline
\textbf{Trace length} & \textbf{EWAH64} & \textbf{Roaring} & \textbf{Concise} & \textbf{Raw} & \textbf{BeepBeep} & \textbf{EWAH32} & \textbf{WAH}\\
\hline\hline
{\href{T96.0.0}{1000000}} & {\href{T96.0.1}{2604166.8}} & {\href{T96.0.2}{2754821.0}} & {\href{T96.0.3}{3030303.0}} & {\href{T96.0.4}{3300330.0}} & {\href{T96.0.5}{947867.3}} & {\href{T96.0.6}{2801120.5}} & {\href{T96.0.7}{2865329.5}}\\

\hline
\end{tabular}
\end{lrbox}

% ----------------------
% Table: tThroughputLongestDFW
% ProcessingtimebyformulaFormulaDecTracelengthbaaaaaaStringlengthb
% ----------------------
\newsavebox{\tThroughputLongestDFW}
\begin{lrbox}{\tThroughputLongestDFW}
\begin{tabular}{|c|c|c|c|c|c|c|c|}
\hline
\textbf{Trace length} & \textbf{EWAH64} & \textbf{Roaring} & \textbf{Concise} & \textbf{Raw} & \textbf{BeepBeep} & \textbf{EWAH32} & \textbf{WAH}\\
\hline\hline
{\href{T98.0.0}{1000000}} & {\href{T98.0.1}{2202643.2}} & {\href{T98.0.2}{2267573.8}} & {\href{T98.0.3}{2398081.5}} & {\href{T98.0.4}{2702702.8}} & {\href{T98.0.5}{757575.75}} & {\href{T98.0.6}{2247191.0}} & {\href{T98.0.7}{2386634.8}}\\

\hline
\end{tabular}
\end{lrbox}

% ----------------------
% Table: tThroughputLongestDZO
% ProcessingtimebyformulaFormulaDabTracelengthbaaaaaaStringlengthb
% ----------------------
\newsavebox{\tThroughputLongestDZO}
\begin{lrbox}{\tThroughputLongestDZO}
\begin{tabular}{|c|c|c|c|c|c|c|c|}
\hline
\textbf{Trace length} & \textbf{EWAH64} & \textbf{Roaring} & \textbf{Concise} & \textbf{Raw} & \textbf{BeepBeep} & \textbf{EWAH32} & \textbf{WAH}\\
\hline\hline
{\href{T100.0.0}{1000000}} & {\href{T100.0.1}{2610966.0}} & {\href{T100.0.2}{2915452.0}} & {\href{T100.0.3}{3021148.0}} & {\href{T100.0.4}{3289473.8}} & {\href{T100.0.5}{2638522.5}} & {\href{T100.0.6}{2747252.8}} & {\href{T100.0.7}{2824858.8}}\\

\hline
\end{tabular}
\end{lrbox}

% ----------------------
% Table: tThroughputLongestDZR
% ProcessingtimebyformulaFormulaDadTracelengthbaaaaaaStringlengthb
% ----------------------
\newsavebox{\tThroughputLongestDZR}
\begin{lrbox}{\tThroughputLongestDZR}
\begin{tabular}{|c|c|c|c|c|c|c|c|}
\hline
\textbf{Trace length} & \textbf{EWAH64} & \textbf{Roaring} & \textbf{Concise} & \textbf{Raw} & \textbf{BeepBeep} & \textbf{EWAH32} & \textbf{WAH}\\
\hline\hline
{\href{T102.0.0}{1000000}} & {\href{T102.0.1}{2754821.0}} & {\href{T102.0.2}{3086419.8}} & {\href{T102.0.3}{2808988.8}} & {\href{T102.0.4}{3401360.5}} & {\href{T102.0.5}{1552795.0}} & {\href{T102.0.6}{2652520.0}} & {\href{T102.0.7}{2832861.2}}\\

\hline
\end{tabular}
\end{lrbox}

% ----------------------
% Table: tThroughputLongestDZW
% ProcessingtimebyformulaFormulaDacTracelengthbaaaaaaStringlengthb
% ----------------------
\newsavebox{\tThroughputLongestDZW}
\begin{lrbox}{\tThroughputLongestDZW}
\begin{tabular}{|c|c|c|c|c|c|c|c|}
\hline
\textbf{Trace length} & \textbf{EWAH64} & \textbf{Roaring} & \textbf{Concise} & \textbf{Raw} & \textbf{BeepBeep} & \textbf{EWAH32} & \textbf{WAH}\\
\hline\hline
{\href{T104.0.0}{1000000}} & {\href{T104.0.1}{2232142.8}} & {\href{T104.0.2}{2444987.8}} & {\href{T104.0.3}{2421307.5}} & {\href{T104.0.4}{2832861.2}} & {\href{T104.0.5}{1390820.6}} & {\href{T104.0.6}{2188183.8}} & {\href{T104.0.7}{2309468.8}}\\

\hline
\end{tabular}
\end{lrbox}

% ----------------------
% Table: tThroughputLongestDZV
% ProcessingtimebyformulaFormulaDafTracelengthbaaaaaaStringlengthb
% ----------------------
\newsavebox{\tThroughputLongestDZV}
\begin{lrbox}{\tThroughputLongestDZV}
\begin{tabular}{|c|c|c|c|c|c|c|c|}
\hline
\textbf{Trace length} & \textbf{EWAH64} & \textbf{Roaring} & \textbf{Concise} & \textbf{Raw} & \textbf{BeepBeep} & \textbf{EWAH32} & \textbf{WAH}\\
\hline\hline
{\href{T106.0.0}{1000000}} & {\href{T106.0.1}{2066115.8}} & {\href{T106.0.2}{2364066.2}} & {\href{T106.0.3}{2450980.5}} & {\href{T106.0.4}{2762431.0}} & {\href{T106.0.5}{1362397.9}} & {\href{T106.0.6}{2252252.2}} & {\href{T106.0.7}{2380952.5}}\\

\hline
\end{tabular}
\end{lrbox}

% ----------------------
% Table: tThroughputLongestDZF
% ProcessingtimebyformulaFormulaDaeTracelengthbaaaaaaStringlengthb
% ----------------------
\newsavebox{\tThroughputLongestDZF}
\begin{lrbox}{\tThroughputLongestDZF}
\begin{tabular}{|c|c|c|c|c|c|c|c|}
\hline
\textbf{Trace length} & \textbf{EWAH64} & \textbf{Roaring} & \textbf{Concise} & \textbf{Raw} & \textbf{BeepBeep} & \textbf{EWAH32} & \textbf{WAH}\\
\hline\hline
{\href{T108.0.0}{1000000}} & {\href{T108.0.1}{2222222.2}} & {\href{T108.0.2}{2364066.2}} & {\href{T108.0.3}{2487562.2}} & {\href{T108.0.4}{2631579.0}} & {\href{T108.0.5}{1388888.9}} & {\href{T108.0.6}{2150537.8}} & {\href{T108.0.7}{2277904.2}}\\

\hline
\end{tabular}
\end{lrbox}

% ----------------------
% Table: tThroughputLongestDZS
% ProcessingtimebyformulaFormulaDahTracelengthbaaaaaaStringlengthb
% ----------------------
\newsavebox{\tThroughputLongestDZS}
\begin{lrbox}{\tThroughputLongestDZS}
\begin{tabular}{|c|c|c|c|c|c|c|c|}
\hline
\textbf{Trace length} & \textbf{EWAH64} & \textbf{Roaring} & \textbf{Concise} & \textbf{Raw} & \textbf{BeepBeep} & \textbf{EWAH32} & \textbf{WAH}\\
\hline\hline
{\href{T110.0.0}{1000000}} & {\href{T110.0.1}{2762431.0}} & {\href{T110.0.2}{2762431.0}} & {\href{T110.0.3}{2777777.8}} & {\href{T110.0.4}{3311258.2}} & {\href{T110.0.5}{1179245.2}} & {\href{T110.0.6}{2770083.0}} & {\href{T110.0.7}{3003003.0}}\\

\hline
\end{tabular}
\end{lrbox}

% ----------------------
% Table: tThroughputLongestDZX
% ProcessingtimebyformulaFormulaDagTracelengthbaaaaaaStringlengthb
% ----------------------
\newsavebox{\tThroughputLongestDZX}
\begin{lrbox}{\tThroughputLongestDZX}
\begin{tabular}{|c|c|c|c|c|c|c|c|}
\hline
\textbf{Trace length} & \textbf{EWAH64} & \textbf{Roaring} & \textbf{Concise} & \textbf{Raw} & \textbf{BeepBeep} & \textbf{EWAH32} & \textbf{WAH}\\
\hline\hline
{\href{T112.0.0}{1000000}} & {\href{T112.0.1}{2754821.0}} & {\href{T112.0.2}{2754821.0}} & {\href{T112.0.3}{2801120.5}} & {\href{T112.0.4}{3355704.8}} & {\href{T112.0.5}{2724795.8}} & {\href{T112.0.6}{2638522.5}} & {\href{T112.0.7}{2732240.5}}\\

\hline
\end{tabular}
\end{lrbox}

% ----------------------
% Table: tThroughputLongestDZN
% ProcessingtimebyformulaFormulaDajTracelengthbaaaaaaStringlengthb
% ----------------------
\newsavebox{\tThroughputLongestDZN}
\begin{lrbox}{\tThroughputLongestDZN}
\begin{tabular}{|c|c|c|c|c|c|c|c|}
\hline
\textbf{Trace length} & \textbf{EWAH64} & \textbf{Roaring} & \textbf{Concise} & \textbf{Raw} & \textbf{BeepBeep} & \textbf{EWAH32} & \textbf{WAH}\\
\hline\hline
{\href{T114.0.0}{1000000}} & {\href{T114.0.1}{2590673.5}} & {\href{T114.0.2}{2793296.0}} & {\href{T114.0.3}{2053388.1}} & {\href{T114.0.4}{2159827.2}} & {\href{T114.0.5}{937207.1}} & {\href{T114.0.6}{1639344.2}} & {\href{T114.0.7}{1893939.4}}\\

\hline
\end{tabular}%
\end{lrbox}

%% ------------------
%% Section: intro
%% ------------------
\section{Introduction}\label{sec:intro} %% {{{

Event traces are produced by computer systems of various kinds: web servers, transactional database systems, and even standard programs that include logging instructions, or that have been instrumented in a special way. Oftentimes, the validity of these traces must be assessed by evaluating a number constraints on the events they contain. For example, one may want to check that each log entry follows a particular format, or that a specific Boolean condition holds for all the events of the trace. Evaluating constraints over a prerecorded trace produced by a system is sometimes called \emph{offline monitoring} --in contrast to online monitoring, which evaluates these conditions on-the-fly as the events are generated. Monitoring, either online or offline, has found applications in various fields, and has proved to be an efficient software testing tool \cite{DBLP:journals/jlp/LeuckerS09,DBLP:journals/sttt/MeredithJGCR12}.

As we shall see in Section \ref{sec:ltl-monitoring}, a pressing problem in monitoring is the evaluation of so-called \emph{stateful} constraints. In contrast with single-event conditions, stateful constraints are related to the ordering of events in the trace. Runtime monitoring is especially focused on the development of efficient algorithms to evaluate such properties, which are generally expressed in formal notations such as $\mu$-calculus, finite-state automata, or temporal logics. However, although \emph{online} monitoring has been the object of much attention, the offline problem has generated less interest. Typically, offline monitoring is seen as the application of online algorithms to a prerecorded event trace. This discards the fact that offline monitoring generally has complete \emph{random} access to the contents of the trace, a fact that could be leveraged to develop more efficient algorithms than for the online case. Moreover, because of the sequential nature of stateful properties, parallelizing their evaluation is a delicate operation that has produced mixed results so far \cite{jocasa,DBLP:journals/fmsd/BasinCEHKM16}.

%\todosylvain{Reformuler l'intro pour commencer par motiver l'usage de LTL pour détecter des patterns dans des logs. Dire que ça peut servir au test d'applications, mais aussi d'autres choses (attaques dans trafic réseau, comportements malicieux, etc.).

%Ensuite, dire: en mode online, beaucoup de travail. En mode offline, peu de choses. La plupart des moniteurs considèrent le offline comme un online à partir d'une trace pré-enregistrée. Presque aucune travail n'exploite le fait que la trace est entièrement disponible et n'a pas à être parcourue linéairement. À l'ère du big data et du parallélisme, peu de travail en analyse de logs

%Reprendre argument de RV12: LTL est séquentiel, comment peut-il être parallélisable? On propose les bitmaps.

%\emph{Temporal logic} \cite{huth2004} is a logistic system which uses rules and symbols to describe and reason about the change of a system's state in terms of time. It is based on the idea that one state may not be constantly true or false as time goes. \emph{Linear Temporal Logic (LTL)} \cite{pnueli77} is a temporal logic, and as its name entails, \emph{LTL} can denote only one sequence of states and for each state there is only one future state.

In this paper, we explore the idea of using bitmap manipulations for the offline evaluation of LTL formul\ae{} on an event log. A \emph{bitmap}, also known as a bit array or bitset, is a compact data structure storing a sequence of binary values. It can be used to express a set of numbers, or an array where each bit represents a 2-valued option. Bitmaps present several advantages as a data structure: they can concisely represent information, and provide very efficient functions to manipulate them, taking advantage of the fact that multiple bits can be processed in parallel in a single CPU instruction. In Section \ref{sec:ltlbitmap}, we introduce a solution which, for a given event trace $\sigma$ and an LTL formula $\varphi$, first converts each ground term into as many bitmaps; intuitively, the bitmap for atomic proposition $p$ describes which events of $\sigma$ satisfy $p$. Algorithms are then detailed for each LTL operator, taking bitmaps as their input and returning a bitmap as their output. The recursive application of these algorithms can be used to evaluate any LTL formula.

This solution presents several advantages. First, the use of bitmaps can be seen as a form of \emph{indexing} (in the database sense of the term) of a trace's content. Rather than being an online algorithm merely reading a prerecorded trace, our solution exploits the fact that the trace is completely known in advance, and makes extensive use of this index to jump to specific locations in the trace to speed up its process. Second, a bitmap having consecutive 0s or 1s can be compressed, which reduces the space cost and speeds up the execution of many operations even further \cite{lemire2014}.

To this end, Section \ref{sec:experiments} describes an experimental setup used to test our solution. It reveals that, that, for complex LTL formul\ae{} containing close to 20 temporal operators and connectives, large event traces can be evaluated at a throughput ranging in the of millions of events per second. This provides a speedup of up to \maxspeedup{}$\times$ compared to a state-of-the art monitor for LTL formul\ae{}. These experiments show that the bitmaps are a compact and fast data structure, and are particularly appropriate for the kind of manipulations required for offline monitoring.

%% }}} --- Section

%% ------------------
%% Section: LTL
%% ------------------
\section{Offline Monitoring for Linear Temporal Logic}\label{sec:ltl-monitoring} %% {{{

Runtime monitoring is the process of analyzing a stream of events produced by the execution of a system. A \emph{monitor} is generally given a specification expressed in some formal language, and describing a property that should hold for all possible executions. In this section, we provide a brief introduction to the concept of monitoring, and to the expression of properties using Linear Temporal Logic. An extensive review of runtime verification is out of the scope of this paper. The reader is referred to introductory material for more details \cite{DBLP:journals/jlp/LeuckerS09,DBLP:series/lncs/BartocciFFR18}.

\subsection{An Overview of Monitoring}

Depending on the context, a monitored system can be instrumented in various ways to report events,  which are caught by the monitor. Monitoring has been successfully applied to a wide range of use cases, including network intrusion detection, object lifecycle verification, bug detection in video games, and security breaches in mobile devices.

Monitoring distinguishes between two modes of operation. In \emph{online} mode, input events are consumed by the system as they are produced, and output events are progressively computed and made available. %It is generally assumed that the output stream is monotonic: once an output event is produced, it cannot be ``taken back'' at a later time.
In contrast, in \emph{offline} mode, the contents of the input streams are completely known in advance (for example, by being stored as a log file or in a database). Whether a system operates online or offline sometimes matters: for example, offline computation may take advantage of the fact that events from the input streams may be indexed, rewound or fast-forwarded on demand. It is precisely this last feature that we shall exploit later in this paper.

When events are recorded to a persistent medium (e.g.\ a file), a trace can be assimilated to a particular form of log. Such a log, in principle, could be parsed and processed by classical command-line tools, such as Grep, or by commercial log analysis systems, such as Snare\footnote{\url{https://www.snaresolutions.com/}},
ManageEngine\footnote{\url{https://www.manageengine.com/products/eventlog/}} or Splunk\footnote{\url{http://www.splunk.com}}. The main difference lies in the fact that typical monitoring properties are \emph{stateful}: the fact that a trace fulfills or violates a given property depends on the ordering of the events in the trace. There exists a large variety of input specification languages for monitors, many of which are based on either finite-state automata, $\mu$-calculus, temporal logic, or variants thereof. In contrast, log analysis software generally lack the means to express complex sequential relationships between elements of a log, and to verify whether such relationships hold for a given log.

\subsection{Linear Temporal Logic}\label{subsec:ltl}

We shall now recall some formal background about one particular specification language, called Linear Temporal Logic (LTL) \cite{pnueli77}.
%
%\subsubsection{Syntax and Semantics}
%
Let $S = \{s_0, s_1, \dots, s_n\}$ be a finite set of symbols called \emph{propositional variables}. An \emph{event} is a total function $e : S \rightarrow \{\top,\bot\}$ that assigns to each propositional variable either the value true ($\top$) or false ($\bot$). A \emph{trace}, noted $\overline{e} = e_0, e_1, \dots, e_n$ is a potentially infinite sequence of events.

LTL formul\ae{} are made of a finite set of atomic propositions, constituting the ground terms of any expression. These propositions can be combined using the Boolean connectives $\neg$, $\wedge$, $\vee$, $\rightarrow$ and temporal logic operators \textbf{F} (eventually), \textbf{G} (globally), \textbf{X} (next), and \textbf{U} (until).

Boolean connectives have their usual meaning. The temporal operator {\bf G}
means ``globally''. For example, the formula $\mbox{\bf G}\,\varphi$ means that
formula $\varphi$ is true in every event of the trace, starting from the
current event. The operator {\bf F} means ``eventually''; the formula
$\mbox{\bf F}\,\varphi$ is true if $\varphi$ holds for some future event of
the trace. The operator {\bf X} means ``next''; it is true whenever $\varphi$
holds in the next event of the trace. Finally, the {\bf U} operator means
``until''; the formula $\varphi\,\mbox{\bf U}\,\psi$ is true if $\varphi$ holds
for all events until some event satisfies $\psi$.
The trace $\overline{e}$ is said to satisfy an LTL formula $\varphi$ if the rules described in Table \ref{tab:semantics} apply recursively. We assume a finite-trace semantics where, if $\overline{e}$ is the empty trace,  $\overline{e} \not\models \tF \varphi$, $\overline{e} \not\models \tX \varphi$, $\overline{e} \not\models \varphi \tU \psi$, but $\overline{e} \models \tG \varphi$.

\vskip -0.25in
\begin{table}
\centering
\begin{tabular}{p{2.25in}p{2.25in}}
\begin{eqnarray*}
\overline{e} \models s_i & \iff &  e_0(s_i) = \top\\
\overline{e} \models \neg\psi & \iff &  \overline{e} \not\models \psi\\
\overline{e} \models \psi \wedge \varphi & \iff &  \overline{e} \models \psi \mbox{ and }\overline{e} \models \varphi\\
%\overline{e}^i \models \psi \vee \varphi & \iff &  \overline{e} \models \psi\mbox{ or }\overline{e} \models \varphi\\
%\overline{e} \models \psi \rightarrow \varphi & \iff &  \overline{e} \not\models \varphi \mbox{ or } \overline{e} \models \psi\\
\overline{e} \models \tX \psi & \iff &  \overline{e}^{1} \models \psi
\end{eqnarray*}
&
\begin{eqnarray*}
\overline{e} \models \tG \psi & \iff &  \forall j \geq 0, \overline{e}^j \models \psi\\
\overline{e} \models \tF \psi & \iff &  \exists j \geq 0, \overline{e}^j \models \psi\\
\overline{e} \models \psi \tU \varphi & \iff &  \exists j \geq i, \pi^j \models \varphi\mbox{ and }\\
& & \forall k, i \leq k < j, \pi^k \models \psi\\
\end{eqnarray*}
\end{tabular}
\vskip -0.25in
\caption{The semantics of LTL. Here $\overline{e}^i$ denotes the subtrace of $\overline{e}$ that starts at event $i$.}
\label{tab:semantics}
\end{table}
\vskip -0.25in

LTL is one of the notations that is widely used in the context of offline monitoring and runtime verification. Depending on the context, LTL formul\ae{} can represent security policies, constraints on sequences of method calls in an object-oriented program \cite{DBLP:conf/ecoop/BoddenHL07}, conditions for robot motion planning \cite{DBLP:journals/ijimai/KumarK16}, reward functions for reinforcement learning \cite{DBLP:conf/ijcai/CamachoIKVM19}, among others. %In addition, LTL specifications can also be extracted from natural language expressions \cite{DBLP:conf/snpd/Luo19}.

%\subsubsection{Richer Semantics}

Representing events as a set of Boolean variables may seem restrictive. However, it is possible to lift propositional LTL to richer types of events. First, Boolean ground terms can be replaced by domain-specific Boolean predicates on events of other types. For example, if input events are numbers, LTL ground terms could represent conditions on these numbers, such as $x = 3 \vee x > 10$, where $x$ is a placeholder for the current event. If events are JSON data structures, ground terms could be expressions like \texttt{x.prop[2] = 'foo'}, in which a JSONPath expression gives a condition on the possible value of an element inside the structure. As long as conditions can be evaluated on individual elements and return a Boolean value, propositional variables can be replaced by such conditions, and only the first line of Table \ref{tab:semantics} needs to be adapted for the case.

We can also consider a particular case where a trace contains interleaved sequences of events for multiple ``instances'' of some entity. A classical example is the monitoring of method calls on Java \texttt{Iterator} objects \cite{DBLP:conf/ecoop/BoddenHL07,DBLP:conf/pldi/JinMGR11}. in such a case, we assume that each event contains a data element that associates it to its corresponding instance (typically a unique process identifier of some form). The sub-sequence of events belonging to the same instance is called a \emph{slice}; applying a separate processing to each such sub-sequence is called \emph{slicing} \cite{DBLP:conf/tacas/ChenR09}. In the case of iterators, the following ``quantified'' LTL expression could represent the fact that, for each instance of the class, a call to its \texttt{next} method must be followed by a call to \texttt{has\-Next}:
\(
\forall i : \tG (\textup{next} \rightarrow (\tX \textup{hasNext}))
\).

In this property, the universal quantifier plays a double role: it is used to separate the input trace into multiple sub-traces depending on the value of some parameter $i$ in each event, and it stipulates that the propositional LTL formula it encloses must hold for every such sub-stream. Although not a full-fledged first-order quantification (since the LTL properties remain propositional and cannot refer to the value of $i$), the increase in expressiveness, compared to standard LTL, is sufficient to handle various real-world use cases. Parametric trace slicing has been used outside of LTL, and is supported by tools based on finite-state automata such as MarQ \cite{DBLP:conf/tacas/RegerCR15} and Larva \cite{CPS09larva}.

%% }}} --- Section

%% ------------------
%% Section: algos
%% ------------------
\section{Evaluating LTL formul\ae{} with Bitmaps}\label{sec:ltlbitmap} %% {{{

Since bitmaps have been shown to be very efficient for storing and manipulating encoded sets of integers, in this section we describe a technique for evaluating arbitrary formul\ae{} expressed in Linear Temporal Logic on a given trace of events through bitmap manipulations.

\subsection{Bitmaps and Compression}\label{subsec:compression} %% {{{

A bitmap (or bitset) is a binary array that we can view as an efficient and compact representation of an integer set. Given a bitmap of $n$ bits, the $i$-th bit is set to 1 if the $i$-th integer in the range $[0, n-1]$ exists in the set. 
It was recognized early on that bitmaps could provide efficient ways of manipulating these sets, by virtue of their binary representation. For example, union and intersection between sets of integers can be computed using bitwise operations (OR, AND) on their corresponding bitmaps; in turn, such bitwise operations can be performed very quickly by microprocessors, and even in a single CPU operation for 32 or 64-bit wide chunks, depending on the architecture.

Furthermore, a bitmap can be used to map $n$ chunks of data to $n$ bits. If the size of each chunk is greater than 1, the bitmap can greatly reduce the size of the storage. In addition, with its capacity of exploiting bit-level parallelism in hardware, standard operations on bitmaps can be very efficient. Unsurprisingly, bitmaps have been used in a lot of applications where the space or speed requirements are essential, such as information retrieval \cite{Chan:1998:BID:276305.276336}, databases \cite{burdick2001mafia}, and data mining \cite{Ayres:2002:SPM:775047.775109,Uno:2005:LVC:1133905.1133916}.

A bitmap with a low fraction of bits set to value 1 can be considered \emph{sparse} \cite{lemire2014}. Such a sparse bitmap, stored as is, is a waste of both time and especially space. Consequently, many algorithms have been developed to \emph{compress} these bitmaps; most of them are based on the Run-Length Encoding (RLE) model derived from the BBC compression scheme \cite{antoshenkov1995byte}. In the following, we briefly describe a few of these techniques. %In particular, we detail the WAH \cite{wu2006optimizing}, Concise \cite{colantonio2010} and EWAH \cite{lemire2010} algorithms, because they have well-implemented open source libraries in Java that we will evaluate experimentally later in this paper.

\subsubsection{WAH}

WAH \cite{wu2006optimizing} divides a bitmap of $n$ bits into $\lceil \frac{n}{w-1}\rceil$ words of $w -1$ bits, where $w$ is a convenient word length (for example, 32). WAH distinguishes between two types of words: words made of just $w-1$ ones ($11\dots 1$) or just $w-1$ zeros ($00\dots 0$), are \emph{fill words}, whereas words containing a mix of zeros and ones are \emph{literal words}. Literal words are stored using $w$ bits: the most significant bit is set to zero and the remaining bits store the heterogeneous $w-1$ bits. Sequences of homogeneous fill words (all ones or all zeros) are also stored using $w$ bits: the most significant bit is set to 1, the second most significant bit indicates the bit value of the homogeneous block sequence, while the remaining $w-2$ bits store the run length of the homogeneous block sequence.

\subsubsection{Concise}

Concise \cite{colantonio2010} is a bitmap compression algorithm based on WAH. Compared with WAH, for which the run length is $w-2$ bits, Concise uses $w - 2 - \lceil \log_2 w \rceil$ for the run length and $\lceil \log_2 w \rceil$ bits to store an integer value indicating to flip a bit of a single word of $w-1$ bits. This feature can improve the compression ratio in the worst case.

\subsubsection{EWAH}

EWAH \cite{lemire2010} is also a variant of WAH but it does not use its first bit to indicate the type of the word like WAH and Concise. EWAH rather defines a $w$-bits marker word. The most significant $w/2$ bits of the word are used to store the number of the following fill words (all ones or all zeros) and the rest $w/2$ bits encodes the number of \emph{dirty words}. These words are exactly like the literal words of WAH, but utilize all $w$ bits. 
%
%With respect to WAH and Concise, the structure used for EWAH makes it difficult to recognize a single word in the sequence as a marker word or a dirty word without reading the sequence from the beginning. Hence, apart from exceptional situations, a reverse enumeration of the bits in the sequence is nearly impossible.

\subsubsection{Roaring}

In all the previous models, fast random access to the bits in an arbitrary sequence is relatively difficult. %At the very least, the word that contains the bit to read must be identified, and the position of this word in the stream requires a knowledge of how many literal or fill words are present before. 
Besides the RLE-model algorithms, there exist other bitmap compression models that support fast random access similar to uncompressed bitmaps. One of them is called ``Roaring bitmap'' \cite{lemire2015}. %, which we shall briefly describe.
It has a compact and efficient two-level indexing data structure that splits 32-bit indexes into chunks, each of which stores the 16 most significant bits of a 32-bit integer and points to a specialized container storing the 16 least significant bits. There are two types of containers: a sorted 16-bit integer array for \emph{sparse} chunks, which store at most 4,096 integers, and a bitmap for \emph{dense} chunks that stores $2^{16}$ integers. %This hybrid data structure allows fast random access whereas all RLE-model algorithms mentioned cannot because of the characteristics mentioned earlier. %The time complexities of the LTL operators with \emph{Roaring bitmap} are $O(n)$ because we cannot skip certain bits when enumerating the bitmap.

\subsection{Manipulating Bitmaps to Implement LTL Operators} %% {{{

We are now ready to define a procedure for evaluating arbitrary LTL formul\ae{} with the help of bitmaps. We suppose that a well-designed bitmap data structure implements a number of basic functions. Given bitmaps $a$, $b$, we will note $|a|$ the function that computes the length of $a$. The notation $a \otimes b$ will denote the bitwise logical AND of $a$ and $b$, $a \oplus b$ the bitwise logical OR, and $!a$ its bitwise inverse, as is shown in Table \ref{tbl:bmfuncs}.

\begin{table}
\centering
\scalebox{0.85}{
\begin{tabular}{|p{0.6in}|p{4in}|}
\hline
\textbf{Function} & \textbf{Description} \\
\hline\hline
$|b|$ & gets the size of the bitmap. \\
\hline
%get(bitmap, index) & gets the true/false value from the specific index in the bitmap. % PAS UTILISÉE DANS LES ALGOS \\
% % \hline
% % set(bitmap, index, val) & sets the true/false value to the specific index in the bitmap. % PAS UTILISÉE NON PLUS \\
% \hline
$\textup{and}(b_1, b_2)$ & performs the pairwise conjunction of the bits in $b_1$ and $b_2$. \\
\hline
$\textup{or}(b_1, b_2)$ & performs the pairwise disjunction of the bits in $b_1$ and $b_2$. \\
\hline
$\textup{not}(b)$ & performs the negation of each bit in $b$. \\
\hline
\end{tabular}
}
\caption{Basic bitmap functions.}
\label{tbl:bmfuncs}
\end{table}

These bitmap functions would be enough to evaluate the LTL operators, but in order to optimize our solution and integrate more closely with bitmap compression algorithms shown in Section \ref{subsec:compression}, we need to manipulate the internal data structure of the bitmap and thus introduce seven derivative bitmap functions see Table \ref{tbl:bmhelpers}).

\begin{table}
\centering
\scalebox{0.85}{
\begin{tabular}{|p{1in}|p{3.5in}|}
\hline
\textbf{Function} & \textbf{Description} \\
\hline\hline
$\textup{addMany}(b, v, \ell)$ & adds an $\ell$-bit sequence of the same value $v$ to the end of the bitmap whose size then increases by $\ell$. \\
\hline
$\textup{copyTo}(b_d, b_s, s, \ell)$ & copies the $\ell$-bits sequence from the index $s$ in bitmap $b_s$ to the end of another bitmap $b_s$ whose size then increases by $\ell$. \\
\hline
$\textup{removeFirstBit}(b)$ & removes the first bit of the bitmap, and the size of the bitmap decreases by 1. \\
\hline
$\textup{next}(v, b, s)$ & gets the position of the next occurrence of the bit with value $v$ starting from position $s$, or $-1$ if there is no more. \\
% \hline
% next(1, bitmap, start) & gets the position of the next occurrence of the bit "1" from the inclusive position $start$ of the bitmap, or -1 if there is no more "1". \\
\hline
$\textup{last}(v, b)$ & gets the position of last occurrence of the bit with value $v$ in the bitmap $b$, or $-1$ if $b$ does not have a bit with value $v$. \\
% \hline
% last(1, bitmap) & gets the position of last occurrence of the bit "1" in the bitmap, or -1 if the bitmap does not have a bit "1". \\
\hline
\end{tabular}
}
\caption{Derivative bitmap functions.}
\label{tbl:bmhelpers}
\end{table}

Given a finite sequence of states $(s_0, s_1, ..., s_{n - 1})$ and an LTL formula $\varphi$, the principle is to compute a bitmap $(b_0b_1...b_ib_{i + 1}...b_{n - 1})$ of length $n$, noted $B_\varphi$, whose content is defined follows:

\begin{equation}\label{eq:map}
b_i = \begin{cases}
1 & \text{if $\overline{s}^i \models \varphi$} \\
0 & \text{otherwise}
\end{cases}
\end{equation}

The finite set of atomic propositions constitute the initial bitmaps. These basic bitmaps are created by reading the original trace, and setting bit $i$ of $B_p$ to 1 if the atomic proposition is true at the corresponding state $s_i$, and otherwise 0. One can see that this construction respects Definition \ref{eq:map} in the case of ground terms.

From these initial bitmaps, bitmaps corresponding to increasingly complex formul\ae{} can now be recursively computed. The cases of conjunction, disjunction and negation are easy to deal with, since these connectives have their direct equivalents as bitwise operators. For example, given bitmaps $B_\varphi$ and $B_\psi$, the bitmap $B_{\varphi \wedge \psi}$ can be obtained by computing $B_\varphi \otimes B_\psi$. The remaining propositional connectives can be easily reduced to these three through standard identities. Propositional logic operators have their direct translation into bitwise operators. %The functions $\mathop{not}(), \mathop{and}(), \mathop{or}()$ from the bitmap are enough to the calculation of the formul\ae{} $\neg\psi$, $\psi \wedge \varphi$, $\psi \vee \varphi$. And the formula $\psi \rightarrow \varphi$ can be expanded to $\neg \psi \vee \varphi$.

\begin{comment}
\begin{align*}
B_{\neg \psi} &\mapsto \mathop{not}(B_\psi) \\
B_{\psi \wedge \varphi} &\mapsto \mathop{and}(B_\psi, B_\varphi) \\
B_{\psi \vee \varphi} &\mapsto \mathop{or}(B_\psi, B_\varphi) \\
B_{\psi \rightarrow \varphi} &\mapsto \mathop{or}(\mathop{not}(B_\psi), B_\varphi) \\
\end{align*}
\end{comment}

\begin{figure}
\centering
\subfloat[$\mbox{\bf X}\,\varphi$]{\includegraphics[scale=0.3]{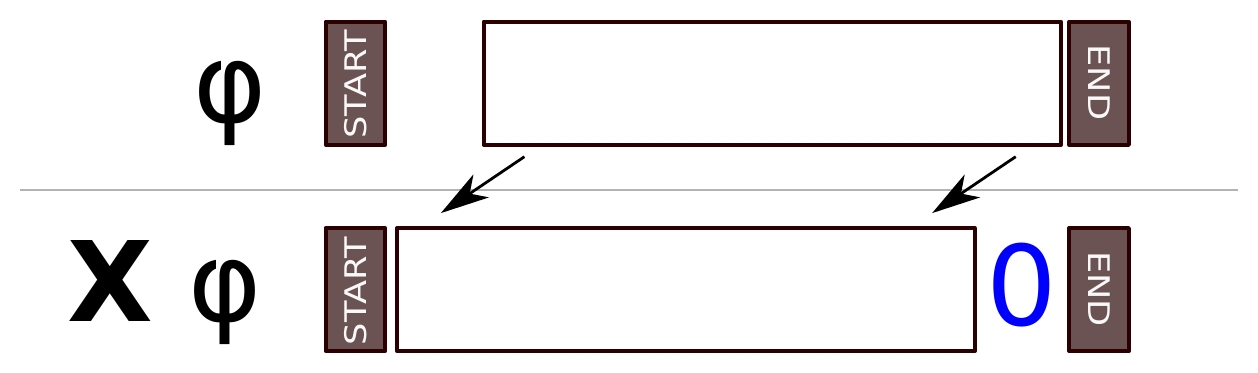}}~~~
\subfloat[$\varphi\,\mbox{\bf U}\,\psi$]{\includegraphics[scale=0.3]{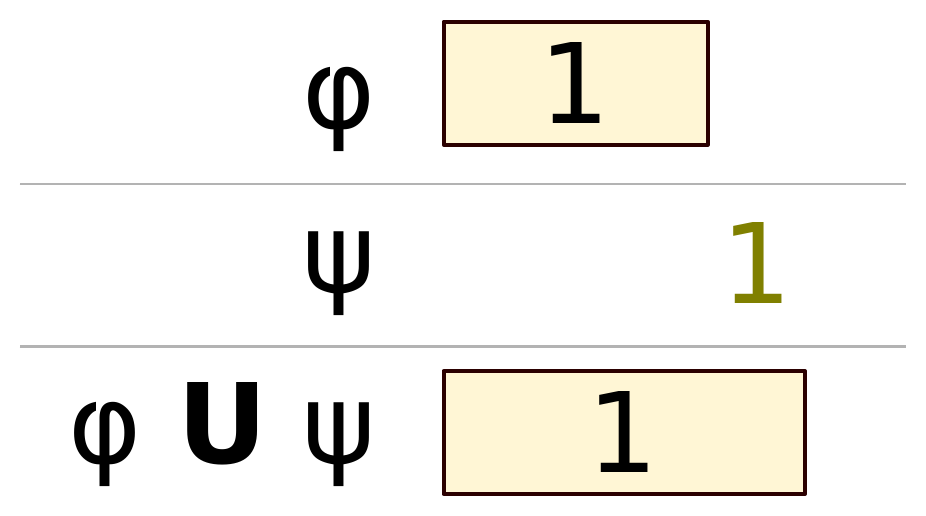}}\\
\subfloat[$\mbox{\bf G}\,\varphi$]{\includegraphics[scale=0.3]{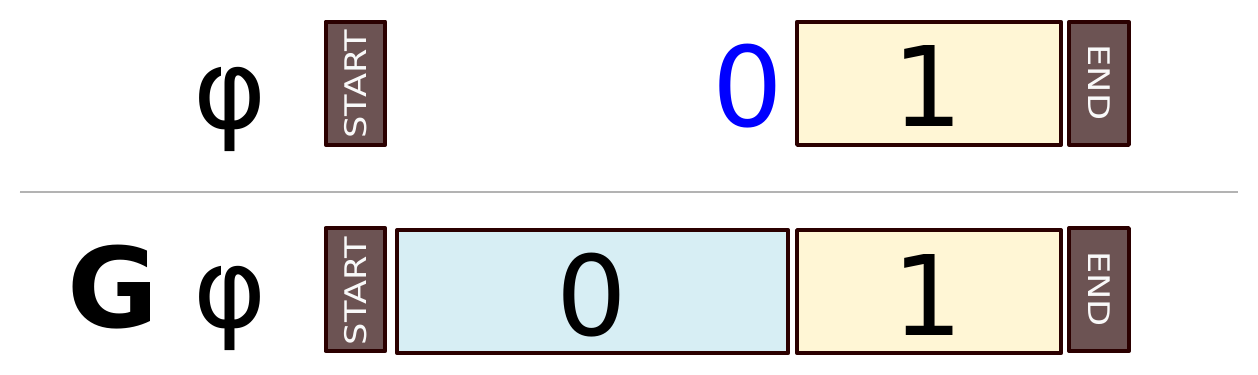}}~~~
\subfloat[$\mbox{\bf F}\,\varphi$]{\includegraphics[scale=0.3]{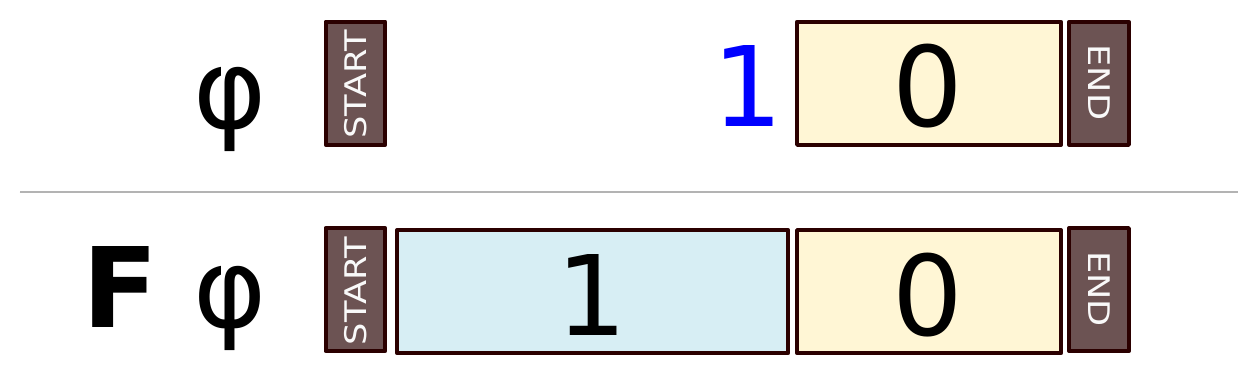}}
\caption{A graphical representation of the computation of four temporal operators on bitmaps}
\label{fig:patterns}
\end{figure}

Temporal logic operators are a little more complicated because they concern the change of the states in terms of time, potentially requiring to enumerate the actual states and the bits in the bitmaps. Still, a few of them can be handled easily. The expression $\tX \varphi$ states that $\varphi$ must hold in the next state of the trace. To compute the bitmap $B_{\tX \varphi}$, it suffices to remove the first state of $B_\varphi$, shift the remaining bits one position to the left, and fill the last bit with 0. This is illustrated in Figure \ref{fig:patterns}a, and formalized in Algorithm \ref{alg:next}.

\begin{algorithm}
\caption{Computing $\tX a$}
\label{alg:next}
\begin{algorithmic}[1]
\Require Bitmap $a$
\State $o \gets$ removeFirstBit($a$)
\State addMany($o$, 0, 1)
\State \Return $o$
\end{algorithmic}
\end{algorithm}

To compute the vector for $\tG \psi$, it suffices to find the smallest position $i$ such that all subsequent bits are 1. In $B_{\tG \psi}$, all bits before $i$ are set to 0, and all bits after (and including) $i$ are set to 1. Thus to implement this operator using bitmaps, we need to do a search in the bitmap $B_{\psi}$ from back to front to find the last occurrence of the bit 0, as can be seen from Algorithm \ref{alg:global}. % and \ref{alg:future}.
Operator \textbf{F} is the dual of \textbf{G}; its corresponding algorithm works in the same way as for \textbf{G}, swapping 0 and 1.

\begin{algorithm}
\caption{Computing $\tG a$}
\label{alg:global}
\begin{algorithmic}[1]
\Require Bitmap $a$
%\State // G(101011) $\Rightarrow$ (000011)
\State $p \gets$ last(0, $a$)
\If {$p = -1$}
  \State \Return $a$
\Else
  \State $o \gets \langle~\rangle$
  %\State $s \gets$ size($a$)
  \State addMany($o$, 0, $p + 1$)
  \State addMany($o$, 1, $|a| - p - 1$)
  \State \Return $o$
\EndIf
\end{algorithmic}
\end{algorithm}

\begin{comment} %% This is the direct dual of G
\begin{algorithm}
\caption{\textbf{F}uture}
\label{alg:future}
\begin{algorithmic}[1]
\Require Bitmap $a$
%\State // F(010100) $\Rightarrow$ (111100)
\State $pos \gets$ last(1, $a$)
\If {$pos = -1$}
  \State \Return $a$
\Else
  \State $o \gets$ empty Bitmap
  %\State $s \gets$ size($a$)
  \State $\textup{addMany}(o, 1, pos + 1)$
  \State addMany($o$, 0, $|a| - pos - 1$)
  \State \Return $o$
\EndIf
\end{algorithmic}
\end{algorithm}
\end{comment}

The last operator to handle is \textbf{U}. According to the formal semantics of LTL, if there is an index $j$ such that $\overline{s}^j \models \psi$ and $\overline{s}^i$ for all $i < j$, then $\overline{s}\models \varphi\tU \psi$. In terms of bitmap operations, we need to keep checking if there is any bit set as 1 in $B_{\varphi}$ before every occurrence of bit 1 in $B_{\psi}$ (see Algorithm \ref{alg:until}).

\begin{algorithm}
\caption{Computing $a \tU b$}
\label{alg:until}
\begin{multicols}{2}
\begin{algorithmic}[1]
\Require Bitmaps $a$ and $b$
\State $o \gets \langle~\rangle$
\State $p, a_0, a_1, b_0, b_1 \gets 0$
%\State $size \gets$ size($a$)
% \State $a_1 \gets 0$
% \State $a_0 \gets 0$
% \State $b_1 \gets 0$
% \State $b_0 \gets 0$
\While {$p < |a|$}
  \If {$a_1 \leq p$}
    \State $a_1 \gets$ next(1, $a$, $p$)
  \EndIf

  \If {$b_1 \leq p$}
    \State $b_1 \gets$ next(1, $b$, $p$)
  \EndIf

  \If {$a_1 = -1$ or $b_1 = -1$}
    \State \textbf{break}
  \EndIf

  \State $n_1 \gets$ min($a_1$, $b_1$)
  \If {$n_1 > p$}
    %\State // (00..) U (00..) $\Rightarrow$ (00..)
    \State addMany($o$, 0, $n_1 - p$)
    \State $p \gets n_1$
    \State \textbf{continue}
  \EndIf

  \If {$p = b_1$}
    %\State // (??..) U (11..) $\Rightarrow$ (11..)
    \If {$b_0 \leq b_1$}
      \State $b_0 \gets$ next(0, $b$, $b_1$)
      \If {$b_0 = -1$}
        \State $b_0 \gets |a|$
      \EndIf
    \EndIf
    \State addMany($o$, 1, $b_0 - p$)
    \State $p \gets b_0$
    \State \textbf{continue}
  \EndIf

  \If {$a_0 \leq a_1$}
    \State $a_0 \gets$ next(0, $a$, $a_1$)
    \If {$a_0 = -1$}
      \State $a_0 \gets |a|$
    \EndIf
  \EndIf
  \If {$a_0 \geq b_1$}
    %\State // (111?..) U (0001..) $\Rightarrow$ (1111..)
    \State addMany($o$, 1, $b_1 - p + 1$)
    \State $p \gets b_1 + 1$
  \Else
    %\State // (11100..) U (00001..) $\Rightarrow$ (00000..)
    \State addMany($o$, 0, $a_0 - p + 1$)
    \State $p \gets a_0 + 1$
  \EndIf
\EndWhile

  \If {$b_1 = -1$}
    %\State // (..??) U (..00) $\Rightarrow$ (..00)
    \State $\textup{addMany}(o, 0, |a| - |o|))$
  \ElsIf {$a_1 = -1$}
    %\State // (..0000) U (..1010) $\Rightarrow$ (..1010)
    \State $\textup{copyTo}(o, b, p, |a| - p)$
  \EndIf

  \State \Return $o$
\end{algorithmic}
\end{multicols}
\end{algorithm}

%% }}} --- Sub-section

\subsection{Discussion} %% {{{

An interesting point of this last algorithm is that the bitmaps $a$ and $b$ are not traversed in a linear fashion. Rather, entire blocks of each bitmap can be skipped to reach directly the next 0 or the next 1, depending on the case. Note that this is only possible if the trace is completely known in advance before starting to evaluate a formula (and moreover, the trace is traversed backwards). Therefore, our proposed solution is an example of an offline monitor that is not simply an online monitor being fed events of a prerecorded trace one by one: it exploits the possibility of \emph{random access} to parts of the trace, which is only possible in an offline setting.

This example shows one of the advantages of our proposed technique in terms of complexity. Indeed, reading the original log to create the ground bitmaps can be done in linear time (and in a single pass for all propositional symbols at once). However, once these initial bitmaps are computed, many of the required operations do not require a linear processing of the trace anymore. For example, evaluating $\tX \varphi$ requires a simple bit shift, which can be done in a single CPU operation for 64 bits at a time, and potentially much more if compression is used.\footnote{The left bit shift of a compressed block is the block itself, as long as the next bit to the right has the same value.} Similarly, looking for the next 0 or 1 seldom requires linear searching, as the use of compression makes it possible to skip over fill words in one operation. Computing the bitmap for a \textbf{F} or \textbf{G} operator requires a single such lookup for the entire trace.

Another interesting point is the fact that operators \textbf{F} and \textbf{G} are monotonous. As can be seen in Figure \ref{fig:patterns}, the resulting bitmap is of the form $0^*1^*$ (or the reverse). Hence a very simple bitmap is propagated upwards to further algorithms; it can be heavily compressed, and makes any lookup for the next 0 or the next 1 trivial. While not producing such simple vectors, bitmaps resulting from the application of \textbf{U} still have a relatively regular structure that is again amenable to reasonable compression.

Although the algorithms we present apply to propositional LTL, they can be lifted to the more expressive LTL variants discussed in Section \ref{subsec:ltl}. First, Boolean conditions on events of a richer type can be evaluated in a single pre-processing pass that replaces each event by a bit vector giving the value of each condition. Ground terms in the LTL expression can then be substituted for the corresponding Boolean variable. Sliced LTL formul\ae{} can also be accommodated with minimal modifications. For each quantifier, it suffices to run a linear pre-processing pass on the original trace that creates one sub-trace for each slice identifier. Assuming, as is often the case, that each event belongs to at most one slice, the sum of the size of each sub-trace is bounded by the length of the original trace. Since evaluating a formula on each of this sub-traces is linear in their length, the global operation remains linear with respect to the input trace.

%% }}} --- Subsection

%% }}} --- Section

%% ------------------
%% Section: expériences
%% ------------------
\section{Implementation and Experiments}\label{sec:experiments} %% {{{

While the worst-case complexity of every algorithm presented in the previous section is still $O(n)$ (where $n$ is the size of the input bitmap), we suspect that performance in practice should be much better. Therefore, in this section, we describe experiments in order to test the performance (both in terms of time and memory) of fundamental LTL algorithms, and their recursive application on complex LTL formul\ae{}.
%\item Evaluate the performance improvements incurred by the use of compression
%\end{enumerate}

\subsection{Experimental Setup} %% {{{

As a means to avoid the runtime disk I/O cost we load all relevant files into memory before the calculations. Thus although using bitmaps can considerably reduce memory requirements, we prepared a workstation with a \machinestring{}. All code is implemented in Java, which takes care of memory management and garbage collection. %To address the delay caused by garbage collection (GC) and especially Full-GC, we called \textit{System.gc()} before and after every formula calculation to provide a runtime environment that was as ``clean'' as possible. 
The JVM used to run the experiments was given \jvmram{} MB of memory.

Table \ref{table:bmlibs} shows the libraries used for different types of bitmaps. In order to implement all the LTL operations, we modified the code of each library to add the necessary functions listed in Table \ref{tbl:bmhelpers} and to optimize the functions so that the time complexity of all operations becomes $O(m)$ (where $m$ is the number of sequences of consecutive 0/1 bits). Roaring bitmaps already provided all the necessary operations and did not need a forked version.

\vskip -0.25in
\begin{table}
\centering
\scalebox{0.85}{
\begin{tabular}{|c|l|}
\hline
Bitmap & Source \\
\hline
Raw & \texttt{java.util.BitSet} from Java SDK \\
\hline
\multirow{2}{*}{WAH} & Original:\ \ \url{https://github.com/metamx/extendedset} \\
& Modified: concealed due to double-blind reviewing %\url{https://github.com/phoenixxie/extendedset}
\\
\hline
\multirow{2}{*}{Concise} & Original:\ \ \url{https://github.com/metamx/extendedset} \\
& Modified: concealed due to double-blind reviewing %\url{https://github.com/phoenixxie/extendedset}
\\
\hline
\multirow{2}{*}{EWAH} & Original:\ \ \url{https://github.com/lemire/javaewah} \\
& Modified: concealed due to double-blind reviewing %\url{https://github.com/phoenixxie/javaewah} 
\\
\hline
Roaring & \url{https://github.com/lemire/RoaringBitmap} \\
\hline
\end{tabular}
}
\caption{The bitmap libraries used in our experimental evaluation.}
\label{table:bmlibs}
\end{table}
\vskip -0.25in

Because of the lack of the support for random access in the RLE-model bitmap compression algorithms, we cannot enumerate the bits in the same way as for an uncompressed bitmap. Therefore we designed an \texttt{Iterator} data structure to store not only the absolute index of current bit in the uncompressed bitmap but also the relative index in the compressed bitmap. Taking the function \texttt{next(1,x)} as an example, if the current relative index is in a sequence word of 0, the search in this word is unnecessary, and we just jump to the next word; if the index is in a sequence word of 1, we return the current index; however, if the index is in a literal word, we have to look for the bit 1 in the $ulen$-bits word.

In order to provide a baseline performance comparison of our proposed approach, we also looked for an existing LTL monitor we could include in the experiments. To provide a fair comparison, the monitor should be written in Java, which excludes approaches implemented in other languages, such as the Maude rewriting engine \cite{DBLP:journals/ase/RosuH05}, or MonPoly \cite{DBLP:conf/rv/BasinKZ17}. We also filter out monitors (both online and offline) whose input specification language is not LTL, which leaves out
 J-Lo \cite{DBLP:journals/entcs/StolzB06}, 
LOLA \cite{DBLP:conf/time/DAngeloSSRFSMM05},
MarQ \cite{DBLP:conf/tacas/RegerCR15}, Mufin \cite{DBLP:conf/tacas/DeckerHS0T16}, PoET \cite{DBLP:conf/sp/ErlingssonS00}, 
PQL \cite{DBLP:conf/oopsla/MartinLL05}, PTQL \cite{DBLP:conf/oopsla/GoldsmithOA05}, 
RuleR \cite{DBLP:journals/logcom/BarringerRH10}, and SEQ.OPEN \cite{DBLP:conf/spin/GaravelM04}. Among the remaining contenders, we find BeepBeep \cite{DBLP:conf/rv/Halle16}, which is an open source event stream processor that is still under active development. A plugin for the tool called \textit{Polyglot} allows a user to write formulas in LTL \cite{DBLP:conf/rv/HalleK18} and to evaluate them against a trace of events.

For the experiments, we developed a random input trace generator. It generates every time a trace with a length ranging between 0 and \maxtracelen{} events. Each event is a tuple of 10 Booleans, corresponding to the values of symbolic state variables $s_0, \dots, s_9$, respectively. The formulas against which these traces are evaluated can be divided in three groups:

\begin{itemize}
\item Simple formulas containing a single LTL operator or Boolean connective are meant to evaluate the ``raw'' performance of each bitmap manipulation algorithm. These formulas are labelled ``Axx'', with \textit{xx} being a numerical identifier.
\item The classical temporal specification patterns introduced by Dwyer \etal{} \cite{DBLP:conf/icse/DwyerAC99}, which can be found on their online repository\footnote{\url{http://patterns.projects.cs.ksu.edu}}. %/documentation/patterns.shtml}}.
The maintainers of the repository argue that these patterns ``occur commonly in the specification of concurrent and reactive systems''. These formulas are labelled ``Dxx'' in the following.
\item A sample of LTL formulas randomly generated by the Spot tool \cite{spot}.  These formulas are labelled ``Sxx''.
\end{itemize}
In total, this brings to \numformulas{} the number of LTL formulas used in our experiments. These formulas contain between 2 and 20 operators, and have a nesting operator depth ranging from 2 to 11. Overall, it can be argued that our selection of LTL formul\ae{} provides a diverse sample, including both expressions occurring in real-world specifications, and convoluted random LTL formulas that can be used as a form of ``stress test''.

The source code, list of LTL formulas and scripts used for this experimental analysis are available online \footnote{URL concealed due to double-blind review. Because of these restrictions, the paper contains a \textbf{temporary} appendix with the most important raw data needed to assess the evaluation.}. Overall, our experiments generated \numdatapoints{} distinct data points.

%% }}}

\subsection{Throughput and Scalability} %% {{{

\begin{figure}
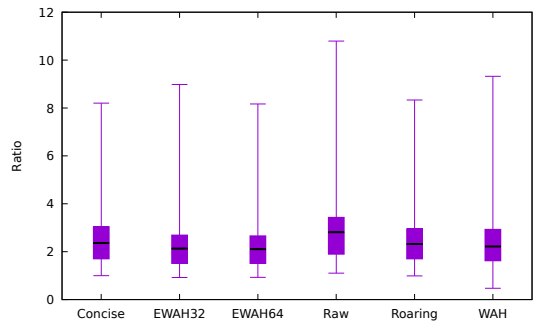

\centering
%\scalebox{0.85}{\usebox{\tThroughputLongestM}}
\scalebox{0.6}{\usebox{\bxRpTime}}
\caption{Relative throughput for each of the bitmap compression methods, with respect to BeepBeep.}
\label{tbl:time}
\end{figure}

A first measurement we took on each formula is its running time, represented here in terms of \emph{throughput} (number of events in the input trace divided by total computation time). Throughput is expressed in Hz (events per second). The results are summarized in Figure \ref{tbl:time}, for input traces of length \maxtracelen{}.

The results are computed as follows: for each input trace and each LTL formula, the throughput $t$ of each bitmap compression method is computed, and compared with the throughput $t'$ of the BeepBeep monitor for the same input. The ratio $t/t'$ indicates by how much a specific bitmap method is faster than BeepBeep for that particular experiment, i.e.\ a value above 1 indicates that the technique outperforms BeepBeep. Figure \ref{tbl:time} shows the boxplot of these ratios, for each bitmap compression method.
These results are unequivocal: bitmap manipulation libraries provide a higher throughput than BeepBeep on all \numformulas{} properties, by a factor ranging between \minspeedup{} and \maxspeedup{}.

Closer examination of the results shows that
%It can be seen from this table that 
the propositional logic operators %(corresponding to A1 and A2)
were faster than most temporal logic operators. Among the temporal logic operators, the binary operators were slower than the unary ones because the former require more operations than the latter, especially in the situation that many 0s and 1s sequences are mixed in the bitmap. The dual operators \textbf{G} and \textbf{F} have similar algorithms but \textbf{F} %(A4)
surprisingly took three times longer than \textbf{G}. %(A5). 
This can be explained by the fact that for a fairly-randomized input bitmap, \textbf{F} will append more 1s than 0s to its output bitmap, while \textbf{G} will append more 0s than 1s. Although the Java \texttt{BitSet} implementation supports both to set a bit to 1 and to clear a bit to 0\footnote{\url{https://docs.oracle.com/javase/8/docs/api/java/util/BitSet.html}}, it actually does nothing when clearing a new bit of which the index is beyond its size, i.e. appending a bit 0. This results in an asymmetrical processing of 0s and 1s in the bitmap.

The results from these first properties suggest that propositional logic operators, temporal logic unary operators and temporal logic binary operators have different magnitudes of processing speed; therefore we can divide these operators into three groups. The remaining experiments %The remaining lines of Table \ref{tbl:time} show the running time for Dwyer \textit{et al.}\@ temporal patterns and the random formul\ae{} generated by Spot.
reveal that the speedup provided by the use of bitmap manipulations is most noticeable for properties having a larger size and/or depth.

\begin{figure}
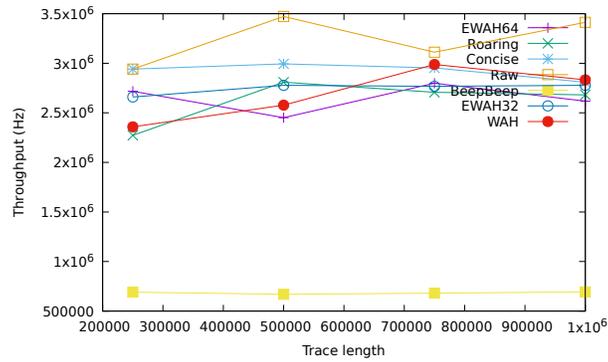

\centering
\scalebox{0.66}{\usebox{\pTimeMFormulaSZOStringPlengthO}}
\caption{Throughput with respect to trace length, for formula S01.}
\label{fig:tp-len}
\end{figure}

A second batch of experiments measured the scalability of each approach ---that is, whether global throughput is influenced by the length of the trace to evaluate. To this end, for each property, we ran each technique on traces of increasing length, ranging from 0 to \maxtracelen{} events. Figure \ref{fig:tp-len} shows a plot resulting from the evaluation of formula S01 (the first of the Spot-generated formul\ae{}) by all techniques. As one can see, there is a roughly constant relationship between the length of the trace and throughput; for bitmap manipulations, this is consistent with our previous observation that all algorithms run in time $O(n)$, with $n$ being the length of the input trace. A linear algorithm indeed implies a constant number of events processed per unit of time. One can also remark that the same applies for the BeepBeep monitor, although with a slower throughput than all bitmap manipulation libraries. Similar trends were observed, at different scales, for all the formul\ae{} in the benchmark.

\begin{figure}
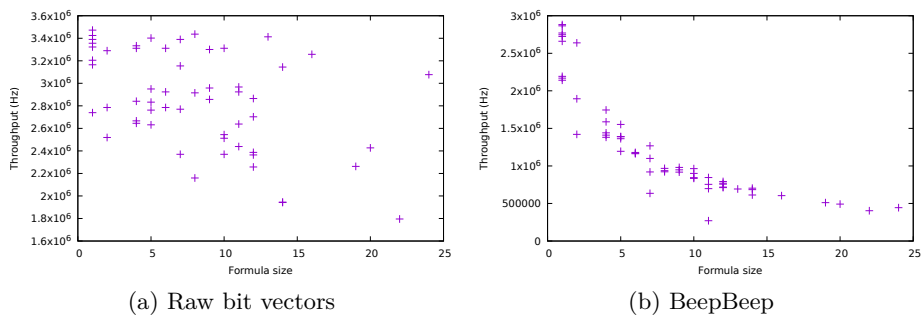

\centering
\subfloat[Raw bit vectors]{\scalebox{0.5}{\usebox{\pThroughputVsSizeLongestRaw}}}~
\subfloat[BeepBeep]{\scalebox{0.5}{\usebox{\pThroughputVsSizeLongestBeepBeep}}}
\caption{Throughput with respect to formula size, for the raw manipulation of bit vectors, and the BeepBeep runtime monitor.}
\label{fig:throughput-size}
\end{figure}

A third batch of experiments focused on the impact of formula size, and whether it impacts negatively on the throughput of all techniques. To perform such a measurement, we plotted the throughput obtained for each formula against its size (in number of operators and Boolean connectives it contains). Figure \ref{fig:throughput-size} shows the results for uncompressed bit vectors, and the BeepBeep monitor, respectively. One can see that the throughput for BeepBeep shows a clear inverse relationship with respect to formula size. This is consistent with the reported implementation of the LTL monitoring algorithm \cite{DBLP:conf/rv/HalleK18}, where each operator handles one input and one output event buffer. The succession of operators entails that a lot of events are passed from one buffer to the next, which results in increased processing time. In contrast, the negative trend for bit vectors is only slightly noticeable. This shows that bit vector manipulations are more robust to handle large LTL formul\ae{}.

%% }}} --- Subsection

\subsection{Resource Usage} %% {{{

A fourth batch of experiments concentrates on memory usage. For these experiments, we counted the maximum number of bytes used up by each technique to evaluate a given LTL formula. Since the algorithm for bitmap manipulation libraries performs a recursive evaluation of operators, the memory it consumes is dominated by the size of all bit vectors that are present on the stack at a given moment. We instrumented our bitmap evaluation routines to measure and store that value. For BeepBeep, memory consumption is dominated by the size of each processor's buffers. We performed a recursive traversal of all \texttt{Processor} objects created by its LTL parser through reflection, recursively counting the amount of memory used by each processor's input and output buffers.

\begin{figure}
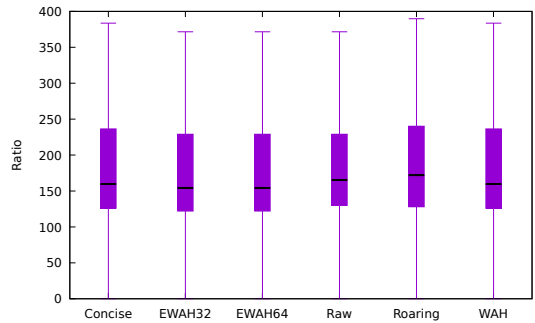

\centering
\scalebox{0.6}{\usebox{\bxRpMem}}
\caption{Relative memory consumption for each of the bitmap compression methods, with respect to BeepBeep.}
\label{tbl:memory}
\end{figure}

The results are summarized in aggregated form in Figure \ref{tbl:memory}, for traces of length \maxtracelen{}. A closer analysis of these measurements has revealed that two situations can occur. This plot has been generated in a similar way as Figure \ref{tbl:time}, using relative memory usage with respect to BeepBeep as the metric. In most cases, bitmap manipulations require slightly more memory than BeepBeep; this is expected, as bitmaps must be completely stored in memory in order to be processed, while BeepBeep processes events one by one and discards them after usage. However, for a few formul\ae{}, such as F02 and F05, BeepBeep's memory usage is, on the contrary, much larger than for bitmap manipulations.

\begin{figure}
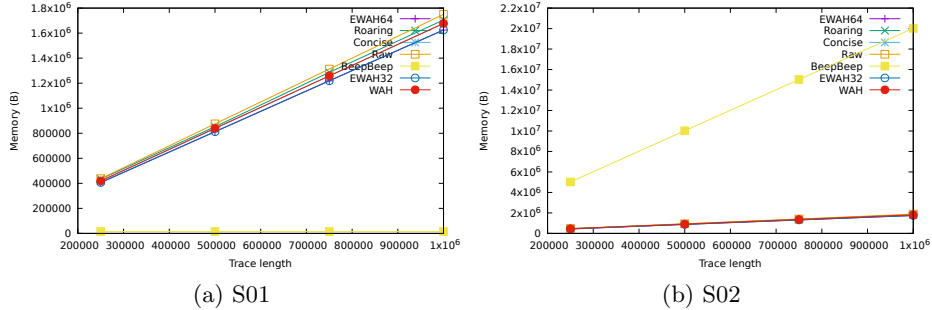

\centering
\subfloat[S01]{\scalebox{0.5}{\usebox{\pMemMFormulaSZOStringPlengthO}}}~
\subfloat[S02]{\scalebox{0.5}{\usebox{\pMemMFormulaSZWStringPlengthO}}}
\caption{Maximum memory consumption with respect to trace length.}
\label{fig:memory-size}
\end{figure}

In order to investigate this behaviour, we plotted memory consumption for each method and each formula, by varying the length of the input trace. Results are shown in Figure \ref{fig:memory-size} for two formul\ae{}. It is important to note that the increasing trace lengths do not represent memory measurements at different moments along the evaluation of the same trace: each data point corresponds to a signe finite trace evaluated from start to end.

We can see that in all cases, memory consumption for bitmap manipulations always increases linearly with the length of the input trace. This is expected, as the bitmaps being manipulated are the same length as the input trace, and the maximum number of such bitmaps present in memory is fixed for a given formula. The shift in BeepBeep's behaviour comes from the fact that, for some instances, its memory consumption becomes \emph{linear} in terms of trace length, as is shown for example in the case of S02.

Further analysis revealed that this situation occurs for LTL formul\ae{} of a particular type. A simple example of such a formula is $(\tF \tG \varphi) \wedge \psi$. By virtue of LTL's semantics, the left member of the conjunction is undefined for any finite prefix of a trace, regardless of $\varphi$; its value can only be decided retrospectively when reaching the end of the trace. If, on the other hand, $\psi$ is an expression that can be evaluated before the end of the trace, the BeepBeep \texttt{Processor} object that handles the conjunction is required to keep the trace of Boolean values produced by $\psi$ in its buffer for the whole duration of the computation --resulting in memory consumption that is ultimately proportional to the length of the input trace. As one can see, when such a situation happens, BeepBeep's memory consumption becomes much larger than for bitmaps.

\begin{comment}
According to the RLE-model algorithms, the compression ratio mostly depends on the length of consecutive 0s or 1s. Hence in this experiment we modified the generator to enable it to repeat the same tuple a specified number of times: 1, 32 and 64. This new mechanism is able to ensure the existence of continuous sequences with a minimum length ($slen$) in the generated bitmaps. Intuitively, when the value of $slen$ increases, the number of sequences decreases; therefore the RLE-model algorithms can be expected to have better performance than an uncompressed bitmap. % for its $O(m)$ time and space complexities.

\begin{figure}
\centering
\usebox{\pTimeTSlenFormulaFZOTracePlengthOZZZZ}
\caption{Throughput with respect to the value of $slen$, for formula F01.}
\label{fig:tp-slen}
\end{figure}

According to Figure \ref{img:f1} and \ref{img:f14}, the performance of the RLE-model algorithms, WAH, EWAH and Concise is obviously related to the value of $slen$. Figure \ref{img:f1} also suggests that the presence of operators \textbf{G} and \textbf{F} can vastly increase the length of consecutive bits of same value, which in turn can be well compressed by RLE-model algorithms. In such a case, several algorithms have better performance than the uncompressed bitmap as $slen$ increases.
\end{comment}

%% }}} --- Subsection

%% }}} --- Section

%% ------------------
%% Section: related work
%% ------------------
\section{Related Work}\label{sec:related} %% {{{

The prospect of using physical properties of hardware to boost the performance of runtime verification has already been studied in the recent past. For example, Pellizzoni \etal\@ \cite{pellizzoni2008hardware} utilized dedicated commercial-off-the-shelf (COTS) hardware \cite{emerson1990temporal} to facilitate the runtime monitoring of critical embedded systems whose properties were expressed in Past-time Temporal Linear Logic (PTLTL). In the past decade, it was shown that the path checking problem (i.e.\ determining if an input trace satisfies an LTL formula) can be parallelized \cite{DBLP:journals/corr/abs-1210-0574}. The process requires the use of a Boolean circuit obtained from unrolling the formula over the trace. This part of the process can be done in parallel for multiple chunks of the input trace at the same time. However, while the evaluation of this unrolling can be done in parallel, a specific type of Boolean circuit requires to be built in advance, which depends on the length of the trace to evaluate.

As the number of cores (GPU or multi-core CPUs) in the commodity hardware keeps increasing, the research of exploiting the available processors or cores to parallelize the tasks and the computing  brings a challenge and also an opportunity to improve the architecture of runtime verification. For example, Ha \etal\@ \cite{ha2009concurrent} introduced a buffering design of \emph{Cache-friendly Asymmetric Buffering} (CAB) to improve the communications between application and runtime monitor by exploiting the shared cache of the multicore architecture; Berkovich \etal\@ \cite{DBLP:journals/fmsd/BerkovichBF15} proposed a GPU-based solution that effectively utilizes the available cores of the GPU, so that the monitor designed and implemented with their method can run in parallel with the target program and evaluate LTL properties.

Closer to our approach is an algorithm introduced by Hallé \etal\@ that uses a cloud computing framework called MapReduce \cite{jocasa}. The algorithm can process multiple, arbitrary fragments of the trace in parallel, and compute its final result through a cycle of runs of MapReduce instances.
%The proposed technique manipulates objects called \emph{tuples}, which are of the form $\langle \phi, (n, i)\rangle$, and are interpreted as the statement ``the process is at iteration $i$, and LTL formula $\phi$ is true for the suffix of the current trace starting at its $n$-th event''. One can see that this statement corresponds exactly to the fact, in the present solution, that the $n$-th position of the bitmap generated by the evaluation of $\phi$ contains the value 1.
%
%Apart from this similarity, however, the two techniques are radically different.
Since the MapReduce approach operates on tuples one by one, while the present solution manipulates entire bitmaps, the algorithms for each LTL operator have little in common with our approach (especially that for \textbf{U}). Where the MapReduce approach gets its speed from the processing of multiple subformul\ae{} on different machines, our present solution is efficient because some operations (such as conjunction) can be computed simultaneously for many adjacent events in a single CPU cycle. In addition, a downside of the MapReduce solution is the large number of tuples generated, and the impossibility of compressing that volume of data.

%\todosylvain{Dire: les bitmaps permettent des opérations en parallèle, mais ce qu'on fait n'a en fait que peu à voir avec le distributed computing.}

As one can see, there have been multiple attempts at leveraging parallelism and properties of hardware to evaluate temporal expressions on traces. However, as far as we know, our work it the first to get its performance boost at the level of the \emph{data structures} used to evaluate these expressions.

%% }}} --- Section

%% ------------------
%% Section: conclusion
%% ------------------
\section{Conclusion and Future Work}\label{sec:conclusion} %% {{{

We proposed a solution for the offline evaluation of LTL formul\ae{} by means of bitmap manipulations. In such a setting, propositional predicates on individual events of a trace states are mapped to bits of a vector (``bitmap'') that are then manipulated to implement each LTL operator. In addition to the fact that bitmap manipulations are in themselves very efficient, our algorithms take advantage of the fact that the trace is completely known in advance, and that random access to any position of that trace makes it possible to skip large blocks of events to speed up the evaluation.

For this reason, our solution is a prime example of an offline evaluation algorithm that exploits the fact that it indeed works offline ---it is not merely an online algorithm that reads events from a prerecorded trace one by one. As a matter of fact, in some cases (such as the \textbf{U} operator), the trace is even evaluated from the end, rather than from the beginning. A thorough performance benchmark for both fundamental operators and complex LTL formul\ae{} proved the feasibility of the approach, and showed how events from a trace can be processed at a rate ranging of millions of events per second.
To further exploit the potential of bitmaps, we introduced bitmap compression algorithms in our solution and integrated them in our benchmark. In the experiments, as we expected, compressed bitmaps demonstrated their ability to easily compress sparse bitmaps and accelerating the LTL operations when there is a certain amount of consecutive bits with the same value. We have explained, how many LTL operators naturally increase the regularity of the bitmaps they are processing.

Obviously, this solution is suitable only for offline evaluation. However, The promising results obtained in our implementation lead to a number of potential extensions and improvements over the current method. First, the algorithm can be reused as a basis for other temporal notations that intersect with LTL, such as Büchi automata. Second, the technique could be expanded to take into account data parameters and quantification. Finally, one could also consider to parallelize the evaluation of large segments of bitmaps on multiple machines.

%% }}} --- Section

%% ---------------------------
%% Bibliography. Uncomment the one corresponding to the
%% stylesheet you want.
%% ---------------------------
%\input{postamble-aaai.inc.tex}
%\input{postamble-acm.inc.tex}
%\input{postamble-acm-journal.inc.tex}
%\input{postamble-elsarticle.inc.tex}
%\input{postamble-ieee.inc.tex}
%\input{postamble-ieee-journal.inc.tex}
%\input{postamble-lipics.inc.tex}
%%%%%%%%%%%%%%%%%%%%%%%%%%%%%%%%%%%%%%%%%%%%%%%%%%%%%%%%%%%%%%%%%%%%%%%%%%
%% This file was autogenerated by PaperShell v1.8 on 2020-05-24 08:42:57
%% https://github.com/sylvainhalle/PaperShell
%% DO NOT EDIT!
%%%%%%%%%%%%%%%%%%%%%%%%%%%%%%%%%%%%%%%%%%%%%%%%%%%%%%%%%%%%%%%%%%%%%%%%%%
\bibliographystyle{splncs03}

%% ---------------------------
%% If there is anything you would like to insert *after* the bibliography
%% (and *before* the \end{document} instruction), place it here
%% ---------------------------

\newpage
\section*{Appendix}

This appendix contains detailed experimental results for throughput and memory consumption, for each LTL formula and each bitmap compression method. %\textbf{This section is not part of the paper} and has been included here for convenience. All these results can also be found in the experimental package that is available online\footnote{URL concealed due to double blind review}.

\subsection*{List of LTL formulas included in the benchmark}

\noindent\textbf{Atomic formulas}
\vskip 0.5cm

\begin{longtable}{ll}
A1 & $\neg (s_0)$\\
A2 & $(s_0) \wedge (s_1)$\\
A3 & $(s_0) \vee (s_1)$\\
A4 & $X (s_0)$\\
A5 & $\tG (s_0)$\\
A6 & $F (s_0)$\\
A7 & $(s_0) U (s_1)$
\end{longtable}

\noindent\textbf{Temporal patterns from Dwyer et al.}
\vskip 0.5cm

\begin{longtable}{lp{5in}}
D01 & $\tG (\neg s_0)$\\
D02 & $((\neg s_0)\tU s_2) \vee (\tG (\neg s_2))$\\
D03 & $\tG ((\neg s_1) \vee (\tG (\neg s_0)))$\\
D04 & $\tG ((\neg s_1) \vee ((\neg s_0)\tU s_2))$\\
D05 & $\tG ((\neg s_1) \vee ((\neg s_0)\tU s_2))$\\
D06 & $F s_0$\\
D07 & $(\tG (\neg s_1)) \vee (F (s_1 \vee (F s_0)))$\\
D08 & $\tG ((\neg s_1) \vee (s_2 \vee ((\neg s_2)\tU(s_0 \wedge (\neg s_2)))))$\\
D09 & $(\neg s_0)\tU(s_0\tU((\neg s_0)\tU(s_0\tU(\tG (\neg s_0)))))$\\
D10 & $(\tG (\neg s_2)) \vee (((\neg s_0) \wedge (\neg s_2))\tU((s_0 \wedge (\neg s_2))\tU(((\neg s_0) \wedge (\neg s_2))\tU((s_0 \wedge (\neg s_2))\tU((\neg s_0)\tU s_2)))))$\\
D11 & $(\neg s_1)\tU(s_1 \wedge ((\neg s_0)\tU(s_0\tU((\neg s_0)\tU(s_0\tU(\tG (\neg s_0)))))))$\\
D14 & $\tG s_0$\\
D15 & $(s_0\tU s_2) \vee (\tG (\neg s_2))$\\
D16 & $\tG ((\neg s_1) \vee (\tG s_0))$\\
D17 & $\tG ((\neg s_1) \vee ((s_0\tU s_2) \vee (\tG (\neg s_2))))$\\
D18 & $\tG ((\neg s_1) \vee (s_0\tU s_2))$\\
D19 & $(\neg s_0)\tU s_3$\\
D20 & $((\neg s_0)\tU(s_2 \vee s_3)) \vee (\tG (\neg s_2))$\\
D21 & $(\tG (\neg s_1)) \vee (F (s_1 \wedge ((\neg s_0)\tU s_3)))$\\
D22 & $\tG ((\neg s_1) \vee (((\neg s_0)\tU(s_2 \vee s_3)) \vee (\tG (\neg s_2))))$\\
D23 & $\tG ((\neg s_1) \vee ((\neg s_0)\tU(s_2 \vee s_3)))$\\
D24 & $\tG ((\neg s_0) \vee (F s_3))$\\
D25 & $(((\neg s_0) \vee ((\neg s_2)\tU((\neg s_2) \wedge s_3)))\tU s_2) \vee (\tG (\neg s_2))$\\
D26 & $\tG ((\neg s_1) \vee (\tG ((\neg s_0) \vee (F s_3))))$\\
D27 & $\tG (((\neg s_1) \vee ((\neg s_0) \vee ((\neg s_2)\tU((\neg s_2) \wedge s_3)))) \vee (\tG (\neg s_2)))$\\
D28 & $\tG ((\neg s_1) \vee (((\neg s_0) \vee ((\neg s_2)\tU((\neg s_2) \wedge s_3)))\tU s_2))$\\
D29 & $((\neg s_0)\tU(s_2 \vee ((\neg s_0) \wedge (s_3 \wedge (X ((\neg s_0)\tU s_4)))))) \vee (\tG (\neg s_2))$\\
D31 & $\tG ((\neg s_1) \vee ((\neg s_0)\tU(s_2 \vee ((\neg s_0) \wedge (s_3 \wedge (X ((\neg s_0)\tU s_4)))))))$\\
D32 & $((\neg s_3)\tU s_0) \vee (\tG ((\neg s_3) \vee (X (\tG (\neg s_4)))))$\\
D33 & $(\tG (\neg s_2)) \vee ((s_2 \vee ((\neg s_3) \vee (X ((\neg s_4)\tU s_2)))) \tU(s_0 \vee s_2))$\\
D34 & $(\neg s_1)\tU(s_1 \wedge (((\neg s_3)\tU s_0) \vee (\tG ((\neg s_3) \vee (X (\tG (\neg s_4)))))))$\\
D37 & $\tG ((\neg s_3) \vee (X ((F (s_4 \wedge (F s_0))) \vee (\tG (\neg s_4)))))$\\
D38 & $\tG ((\neg s_1) \vee (\tG ((\neg s_3) \vee (X ((\neg s_4)\tU(s_4 \wedge (F s_0)))))))$\\
D39 & $\tG ((\neg s_0) \vee (F (s_3 \wedge (X (F s_4)))))$\\
D40 & $\tG ((\neg s_1) \vee (\tG ((\neg s_0) \vee (s_3 \wedge (X (F s_4))))))$\\
D41 & $\tG ((\neg s_0) \vee (F (s_3 \wedge ((\neg s_5) \wedge (X ((\neg s_5)\tU s_4))))))$\\
D42 & $\tG ((\neg s_1) \vee (\tG ((\neg s_0) \vee (s_3 \wedge ((\neg s_5) \wedge (X ((\neg s_5)\tU s_4)))))))$\\
\end{longtable}

\noindent\textbf{Formulas generated by Spot}
\vskip 0.5cm

\begin{longtable}{lp{5in}}
S01 & $(\neg(\tX (\tX ((\tG s_5) \wedge (\tX ((\neg s_0) \vee s_3)))))) \rightarrow ((\tX s_8) \wedge (\tX (s_8 \rightarrow s_4)))$\\
S02 & $((\tX (s_6 \rightarrow s_1)) \wedge ((\neg s_6) \rightarrow ((\neg(\tX (\tX (\tG (\tF (s_1 \wedge s_2)))))) \rightarrow s_5))) \vee (\tF ((\tG (\tF (\tX s_6))) \rightarrow ((\neg(s_2 \vee s_6)) \rightarrow (s_4 \vee (\tG (\neg s_0))))))$\\
S03 & $\tF ((\neg(s_3 \vee s_9))\tU(s_5 \vee s_6))$\\
S04 & $(s_3\tU s_2)\tU s_6$\\
S05 & $s_5 \vee (((\tG (\tF s_6)) \rightarrow (\neg(\tX (\tF s_6))))\tU((\tF s_6) \rightarrow (\tX s_6)))$\\
S06 & $(\neg s_9) \wedge ((\tG s_4) \wedge ((\tX (\tF ((s_7 \vee (\tG s_7)) \wedge (\tF (\tG s_0)))))\tU(\tX (\tX (s_7\tU((s_0 \rightarrow (\neg s_1)) \wedge (s_3 \vee (s_7 \wedge (\neg(\tX s_9))))))))))$\\
S07 & $(\tG (\neg(\tF s_3)))\tU((\tF s_6) \vee ((((\tX s_7) \rightarrow (\tF (\neg s_2))) \vee (\tX (\tX (\tX s_6)))) \rightarrow (\tG s_7)))$\\
S08 & $\tF (s_7 \wedge ((\tG (\tF ((\tX (\neg(s_3\tU s_4))) \vee ((\tG s_8)\tU(\neg s_1))))) \vee (s_3\tU(s_7\tU s_1))))$\\
S09 & $(\neg(\tG ((((\tG (\tF (\tG s_0)))\tU(s_2\tU s_9))\tU(\neg(\tX s_0)))\tU s_7))) \rightarrow (\tF (\neg s_4))$\\
S10 & $(\neg((\tX (\tG s_2))\tU(\tG (\tX (\tF (\tX (\tX s_1))))))) \wedge (s_0 \vee (\tX (\tX (\tX s_3))))$\\
S11 & $\tX ((\neg s_0) \wedge (s_4\tU s_2))$\\
S12 & $(\tX (\tG (\tX (\tX (((\tG (\neg(\tG (s_1 \vee s_4))))\tU(\tF (s_1 \wedge s_6))) \rightarrow (\tX ((\tX (s_0 \wedge s_7))\tU s_4))))))) \rightarrow ((\neg s_3) \wedge (s_5\tU s_2))$\\
S13 & $\tX ((s_9 \wedge (s_1 \vee (\neg s_6))) \rightarrow (s_4 \vee (\neg(\tX s_5))))$\\
\end{longtable}

\begin{figure}
\centering
\scalebox{0.65}{\usebox{\tThroughputLongestM}}
\caption{Throughput, in Hz, for the evaluation of LTL formul\ae{} ($f$). R = Roaring, E32 = EWAH32, E64 = EWAH64, C = Concise, B = BeepBeep.}
\label{tbl:throughput-all}
\end{figure}

\begin{figure}
\centering
\scalebox{0.65}{\usebox{\tMemLongestM}}
\caption{Memory consumption, in millions of bytes, for the evaluation of LTL formul\ae{} ($f$). R = Roaring, E32 = EWAH32, E64 = EWAH64, C = Concise, B = BeepBeep.}
\label{tbl:memory-all}
\end{figure}

%% :wrap=soft:
\end{document}